\documentclass[manuscript]{aastex63}

\received{}
\revised{}
\accepted{November 23, 2019}

\usepackage{soul}

\submitjournal{ApJ}

\begin{document}

\title{\MakeUppercase{Early-type Host Galaxies of Type Ia Supernovae.
II. Evidence for Luminosity Evolution in Supernova Cosmology}}

\correspondingauthor{Young-Wook Lee}
\email{ywlee2@yonsei.ac.kr}

\author[0000-0002-5261-5803]{Yijung Kang}
\affil{Center for Galaxy Evolution Research \& Department of Astronomy, Yonsei University, Seoul 03722, Republic of Korea}

\author[0000-0002-2210-1238]{Young-Wook Lee}
\affil{Center for Galaxy Evolution Research \& Department of Astronomy, Yonsei University, Seoul 03722, Republic of Korea}

\author[0000-0002-1031-0796]{Young-Lo Kim}
\affil{Universit\'e de Lyon, F-69622, Lyon, France; Universit\'e de Lyon 1, Villeurbanne; CNRS/IN2P3, Institut de Physique Nucl\'eaire de Lyon}
\affil{Center for Galaxy Evolution Research \& Department of Astronomy, Yonsei University, Seoul 03722, Republic of Korea}

\author[0000-0001-6812-4542]{Chul Chung}
\affil{Center for Galaxy Evolution Research \& Department of Astronomy, Yonsei University, Seoul 03722, Republic of Korea}

\author[0000-0001-8986-112X]{Chang Hee Ree}
\affil{Korea Astronomy and Space Science Institute, Daejeon 34055, Republic of Korea}

\begin{abstract}
The most direct and strongest evidence for the presence of dark energy is provided by the measurement of galaxy distances using SNe Ia. This result is based on the assumption that the corrected brightness of SN Ia through the empirical standardization would not evolve with look-back time. Recent studies have shown, however, that the standardized brightness of SN Ia is correlated with host morphology, host mass, and local star formation rate (SFR), suggesting a possible correlation with stellar population property. To understand the origin of these correlations, we have continued our spectroscopic observations to cover most of the reported nearby early-type host galaxies. From high-quality (signal-to-noise ratio $\sim$175) spectra, we obtained the most direct and reliable estimates of population age and metallicity for these host galaxies. We find a significant correlation between SN luminosity (after the standardization) and stellar population age at a 99.5\% confidence level. As such, this is the most direct and stringent test ever made for the luminosity evolution of SN Ia. Based on this result, we further show that the previously reported correlations with host morphology, host mass, and local SFR are most likely originated from the difference in population age. This indicates that the light-curve fitters used by the SNe Ia community are not quite capable of correcting for the population age effect, which would inevitably cause a serious systematic bias with look-back time. Notably, taken at face values, most of the Hubble residual used in the discovery of the dark energy appears to be affected by the luminosity evolution. 
\end{abstract}


\section{Introduction} \label{Section1}
Despite the general consensus, the presence and nature of the dark energy are currently the most serious conundrum in astrophysics and cosmology. The distance measurements using SNe Ia for the galaxies at high redshift have provided the most direct evidence for the accelerating universe with dark energy \citep{Riess1998,Schmidt1998,Perlmutter1999}. To use the SNe Ia as a standard candle, however, the ``standardization'' is required because of a large spread in their intrinsic luminosity. This is fulfilled through the empirical procedures based on their light-curve shape and color \citep[e.g.,][]{Phillips1993,Riess1996,Perlmutter1997,Guy2007,Jha2007}. In SN cosmology, it is further assumed that this standardized SN Ia brightness does not evolve with look-back time. However, in observational cosmology, there is always a possibility of the luminosity evolution of standard candle with redshift, as first pointed out by \citet{Tinsley1968}. Obviously, SN progenitors in host galaxies are getting younger with redshift, and several investigators had some doubts on the assumption of no luminosity evolution in SN cosmology although they were not based on observational evidence \citep[e.g.,][]{Drell2000, Ferramacho2009,Linden2009,Tutusaus2017,Tutusaus2019}. According to the analysis by \citet{Childress2014}, the difference in the mean age of SN progenitors between $z$ = 0 and 1.0 is predicted to be $\sim$5.3 Gyr (see also Section~\ref{Section4} below). It is therefore possible that SNe at high redshift are fainter (after the standardization) than the value expected from the model without the dark energy, not because of the dark energy, but mostly because of the luminosity evolution. Careful investigation of this possible systematic bias with redshift is particularly important because the dimming of SNe is only $\sim$0.2 mag ($\sim$20\% in brightness) with respect to a cosmological model without dark energy ($\Omega_{M}$ = 0.27, $\Omega_{\Lambda}$ = 0.00; \citealt{Hicken2009}).

 Nevertheless, to test the effect of luminosity evolution, in the discovery papers, \citet{Riess1998}, \citet{Schmidt1998}, and \citet{Perlmutter1999} only used morphological classification of host galaxies in the local universe as a proxy for stellar population age. Because of apparently very small difference in the standardized brightness between SNe Ia in the early-type and late-type host galaxies, they concluded that the luminosity evolution is negligible in SN cosmology. Later analysis based on a larger sample by \citet{Hicken2009}, however, found a systematic difference of $\sim$0.14 mag between the very early- and very late-type galaxies \citep[see also][]{Suzuki2012}. Recent investigations of host galaxies further found a correlation between the SN brightness\footnote{Throughout this paper, unless otherwise specified, by SN Ia brightness/luminosity, we are referring to the standardized brightness/luminosity after the light-curve shape and color corrections.}and host mass, suggesting that SNe Ia in less massive galaxies are $\sim$0.1 mag fainter than those in more massive galaxies \citep{Kelly2010,Sullivan2010,Childress2013,Johansson2013}. More recent studies based on star formation rate (SFR) in host galaxies also showed that the SNe Ia in locally star-forming environments are fainter than those in locally passive environments \citep{Rigault2015,Rigault2018,Kim2018,Roman2018,Barkhudaryan2019,Rose2019}. As the host morphology, host mass, and the physics of star formation cannot directly affect SN luminosity, all of these correlations are probably not directly originated from these host properties, but more likely due to the age or metallicity of the SN Ia progenitor closely related to these properties.

To directly test this, 9 yr ago, we have initiated the project YONSEI (YOnsei Nearby Supernovae Evolution Investigation). For this project, we have observed $\sim$60 target galaxies selected from the YONSEI SN catalog \citep{Kim2019}, covering most of the reported nearby ($0.01 < z  < 0.08$) early-type host galaxies. The standardized brightnesses of SNe Ia in these target galaxies were obtained from the light-curve fitters implemented in the SuperNova ANAlysis software \citep{Kessler2009} package. Our target galaxies are restricted to the early-type galaxies (ETGs) because a reliable age dating is possible for them from the absorption-line dominated spectra, and their dust extinction is relatively negligible (see \citealt{Gallagher2008} for the pioneering work). For each target galaxy, the absorption lines from very high quality (signal-to-noise ratio (S/N) $\sim$175) spectra have been used to determine population age, metallicity, and velocity dispersion by employing the population synthesis models. This is the well-established technique in the extragalactic community that can provide the best age dating for distant galaxies where the individual stars are not resolved \citep{Faber1992,Worthey1994,Worthey1997,Cardiel1998,Trager2000,Thomas2005,Thomas2010, Graves2009}.

In Paper I of this series \citep{Kang2016}, based on 27 host galaxies, we reported that early-type host galaxies are also following the ``downsizing'' trend, the well-known correlation between the mass (velocity dispersion) and population age of nonhost galaxies. This result suggested that stellar population age is mainly responsible for the correlation between host mass and standardized SN brightness, which, when confirmed, would imply that the luminosity evolution plays a major role in the systematic uncertainty of SN cosmology. Here, as the second paper of this series, we aim to directly confirm the possible luminosity evolution of SN Ia with population age. We have obtained spectra for 32 additional host galaxy sample which has been combined with the dataset previously reported in Paper I, securing a sufficient number of host galaxies required for the purpose of this paper. We have also revisited previously reported correlations of SN brightness with host morphology, host mass, and SFR,to compare them with our result based on population age.

\section{Observations and Data Analyses} \label{Section2}

In Paper I, we reported spectroscopic observations made for the nearby early-type host galaxies at Las Campanas Observatory (LCO) 2.5 m telescope in the southern hemisphere. Here, we have continued our observations at MMT 6.5-m telescope for the targets observable from the northern hemisphere. As in Paper I, our host galaxy sample is limited to nearby targets in the redshift range of 0.01 $\textless$ $ z $ $\textless$ 0.08 with 11.69 ${\textless}$ $B$ ${\textless}$ 18.4 mag. We have selected only ETGs (E-S0) based on the morphological classification in the NASA Extragalactic Database (NED) or the HyperLeda database \citep{Makarov2014}.

The light-curve parameters for SNe occurred in host galaxies were extracted from the YONSEI SN catalog \citep{Kim2019} which consistently combined the data from various surveys over a wide redshift range. This catalog contains the light-curve shape, color and extinction parameters, together with the values for the observed rest-frame peak apparent magnitude in $B$-band ($m_{B}$) and the estimated distance ($\mu_{B}$) calculated from two independent light-curve fitters, SALT2.4 \citep{Guy2007, Betoule2014} and MLCS2k2 \citep{Jha2007}. The best-fit cosmological parameters from this catalog are $\Omega_{M}$ = 0.30, $H_{0}$ = 70, $\alpha$ = 0.15, $\beta$ = 3.69, and $M_{B}$ = -19.06 mag for SALT2, assuming a flat $\Lambda$CDM model. From these parameters, the value for the Hubble residual (HR) is estimated, which is defined as the difference in distance modulus between that measured from SN Ia and that predicted from adopted cosmology (HR $\equiv \mu_{SN} - \mu_{model}(z)$).\footnote{Here we have adopted the $\Lambda$CDM model ($\Omega_{M}$ = 0.30, $\Omega_{\Lambda}$ = 0.70) for the calculation of the HR. However, as we are dealing with only the nearby sample, almost identical values would be obtained even we had adopted other cosmological model.} As the purpose of this paper is to examine the systematic variation of SN luminosity with host property, the intrinsic dispersion ($\sigma_{int}$) is set to 0, as generally adopted in SN host galaxy studies \citep[e.g.,][]{DAndrea2011,Gupta2011,Pan2014}. In this paper, we only use the parameters derived from SALT2 because those from MLCS2k2 have higher intrinsic dispersion likely due to the lack of recent calibration \citep[see][]{Jones2015}. The light-curve parameters listed in the YONSEI catalog are in good agreements with those obtained by other investigators. The readers are referred to \citet{Kim2019} for a more detailed description of this catalog.

\begin{figure}

\epsscale{.9}
\plotone{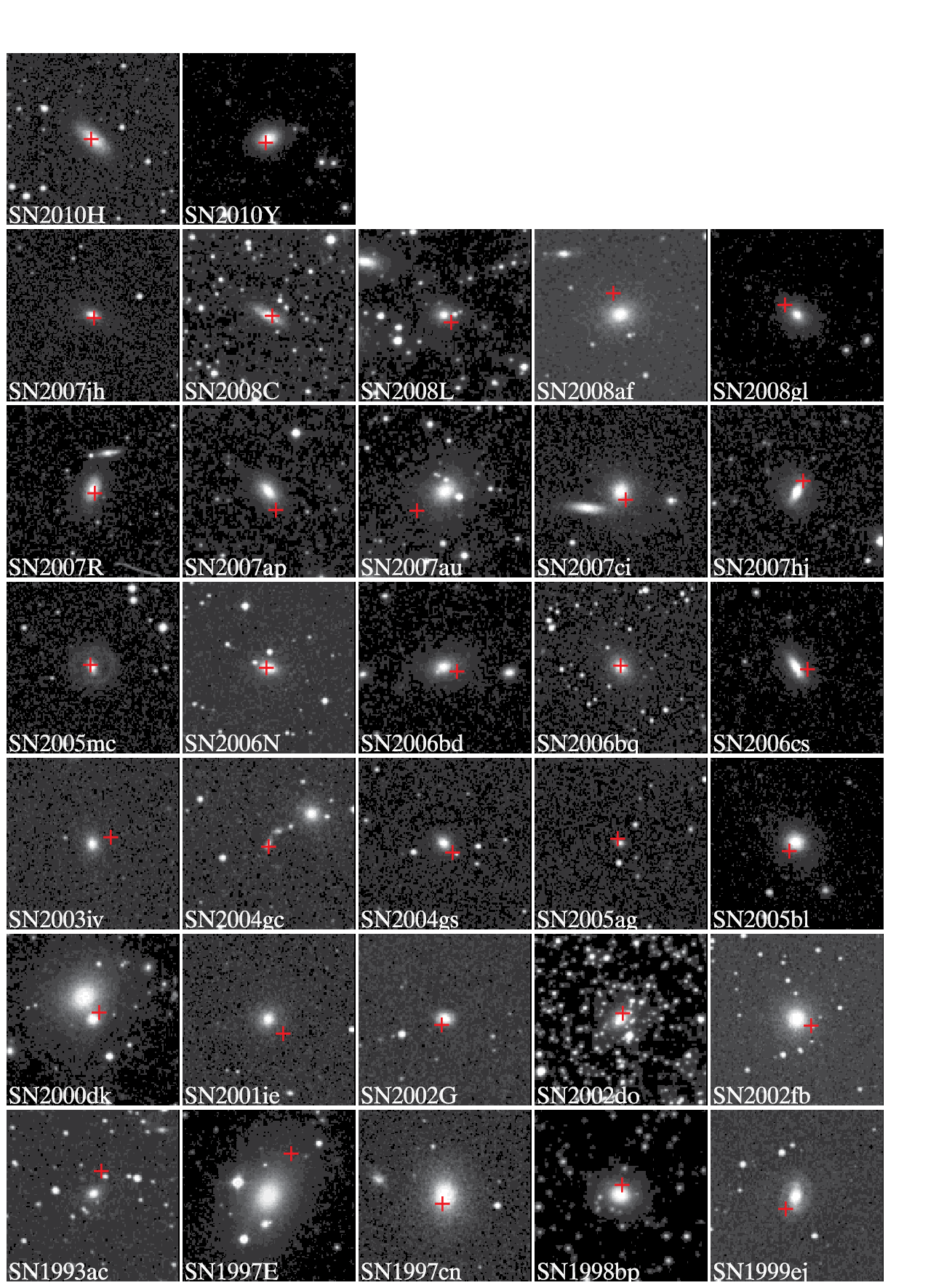}
\caption{\label{figure1} DSS images of our host galaxy sample observed at MMT 6.5 m.  Each of host galaxies is placed at the center of an image, and the name and the position (the red cross) of SN Ia are given in each panel. The images of host galaxies observed at LCO can be found in Figure 1 of Paper I. 
}
\end{figure}

As the purpose of this paper is to investigate the correlation between SN brightness and global host property, additional target galaxies observed at MMT 6.5-m were selected for which SN light-curve parameters, $m_{B}$, and/or $\mu_{B}$ are provided in YONSEI SN catalog. In addition, we only selected host galaxies in which SNe are located within 5 effective radius \citep[as defined by][]{Cappellari2011} so that the population property of SN progenitor would not be very different from the global property measured in the central region of each galaxy.  Among the host galaxy sample of Paper I observed at LCO, 20 galaxies satisfy these criteria. A host galaxy of SN~2007ba, UGC~09798, was further excluded from our sample due to an extreme peculiarity of its SN \citep[see][]{Hicken2009,Burns2014,Scalzo2019}. Table~\ref{table1} lists the basic information of our total sample of host galaxies to be used in this paper, including 19 galaxies among those reported in Paper I. The SALT2 light-curve parameters of SNe Ia in our sample are listed in Table~\ref{table2}. For 32 host galaxies added in this paper, the digitized sky survey (DSS) images marked with SN Ia position are shown in Figure~\ref{figure1}.

For the additional sample of host galaxies, observations have been carried out using the long-slit Blue Channel Spectrograph on MMT 6.5 m during the three observing runs in 2014 May, 2015 January, and 2016 December. As listed in Table~\ref{table1}, the typical single exposure time for each galaxy was 600 - 1800 s depending on the target brightness and/or the sky condition. The settings for the slit position angle (PA) and for the calibration frames (dome flats, twilight sky flats, and HeNeAr arc lamp) are mostly identical to those described in Paper I.  Table~\ref{table3} lists an instrumental setup for our long-slit spectroscopy. Following the usual manner, we have observed standard stars for flux calibration, radial velocity correction, and telluric feature removal. To transform the measured indices to the Lick/IDS standard system, we have further observed 27 stars from the Lick library \citep{Wortheyetal1994,Worthey1997}. To check for the possible systematic offset between the measured indices from LCO and MMT, we have also revisited 11 early-type galaxies (five host galaxies, six nonhost galaxies) covered in Paper I. In addition to this, four more nonhost galaxy spectra were taken for the consistency check with previous studies in the literature.

\begin{figure}
\center
\epsscale{1.2}
\plotone{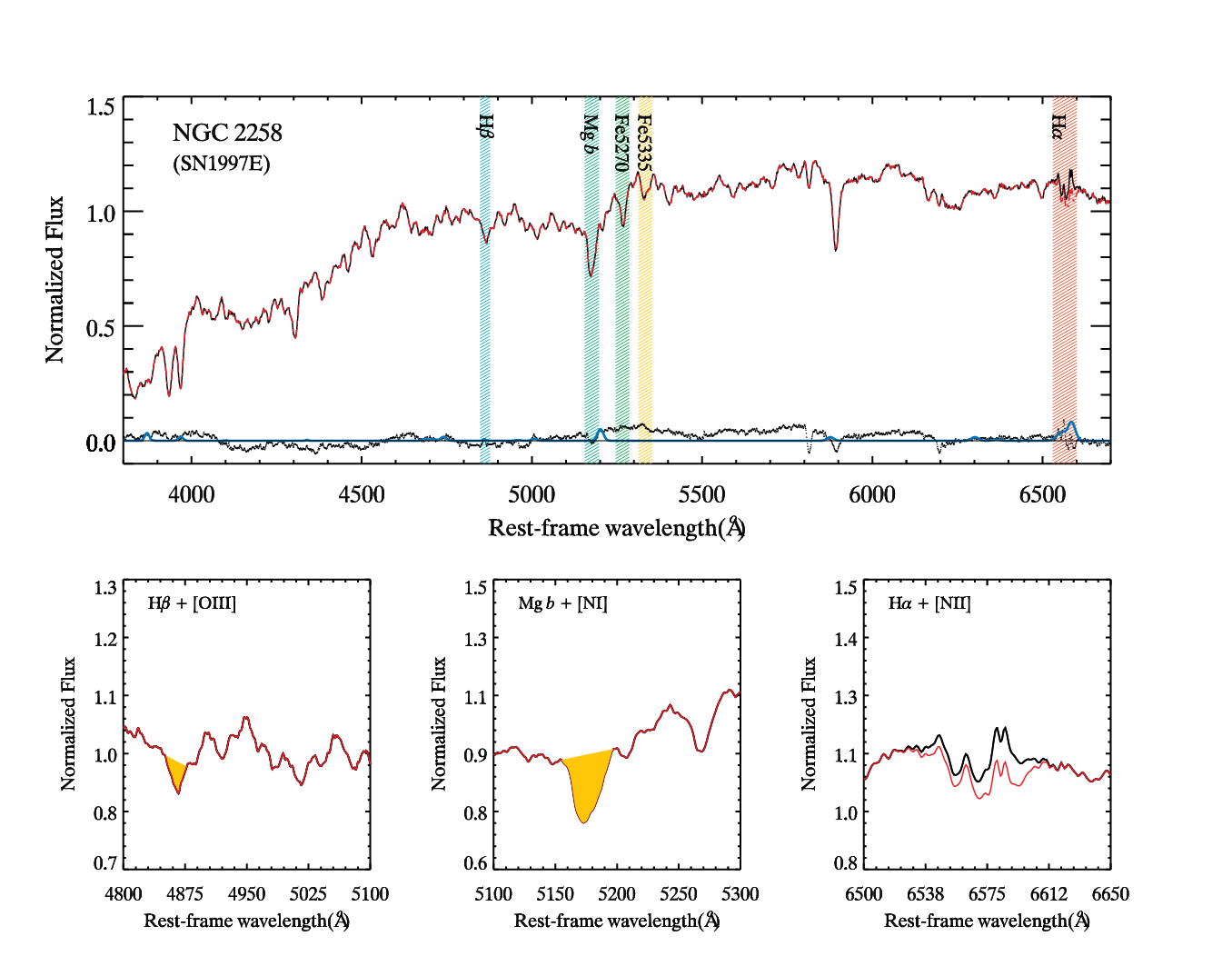}
\caption{\label{figure2}Example of our host galaxy spectra for NGC~2258, a host of SN~1997E. The black and red solid lines are fully calibrated spectra in the rest frame, before and after the emission correction, respectively. The upper panel shows the entire wavelength coverage of our observation, where the absorption bands for H$\beta$, Mg\,b, Fe5270, Fe5335, and H$\alpha$ are indicated in color shades. The gray line at the bottom of the panel shows the difference between the best-fit model and our spectrum, where the detected emission lines (cyan) are overlapped on the residuals. The lower panels show the spectral regions around H$\beta$+[O III] (4900-5100 {\AA}), Mg\,b+[N I] (5100-5300 {\AA}), and H$\alpha$+[N II] (6500-6650 {\AA}), where the orange shades indicate the areas for the line strength measurements}.
\end{figure}

\begin{figure}
\center
\epsscale{0.9}
\plotone{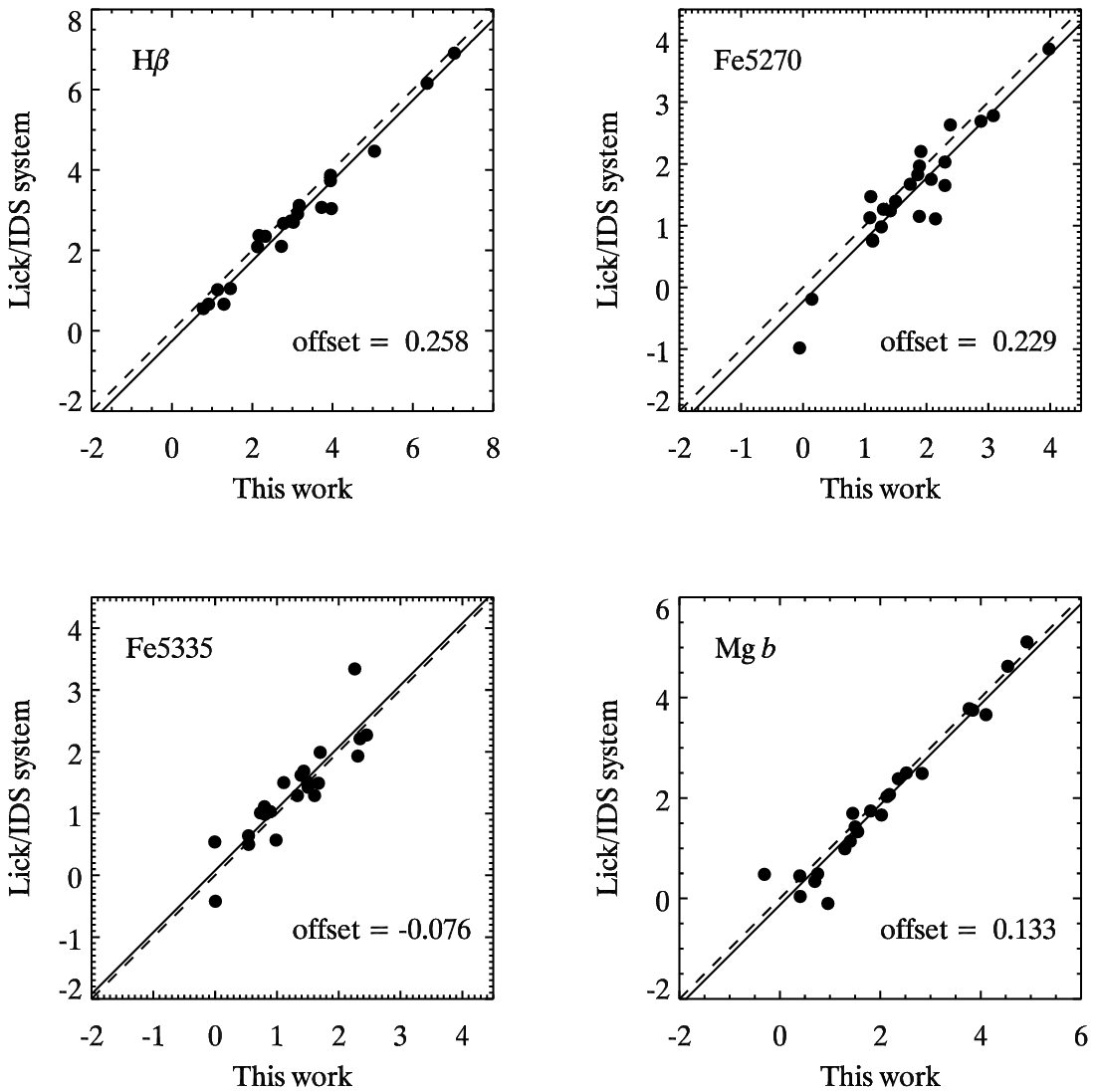}
\caption{\label{figure3}Comparison of Lick indices measured in this study for the Lick/IDS standard stars with those in \citet{Wortheyetal1994} and \citet{Worthey1997}. The dashed line is for the one-to-one relation, while the solid line shows the mean offset (our work-Lick/IDS system) which is given in each panel.}
\end{figure}

\begin{figure}
\center
\epsscale{0.9}
\plotone{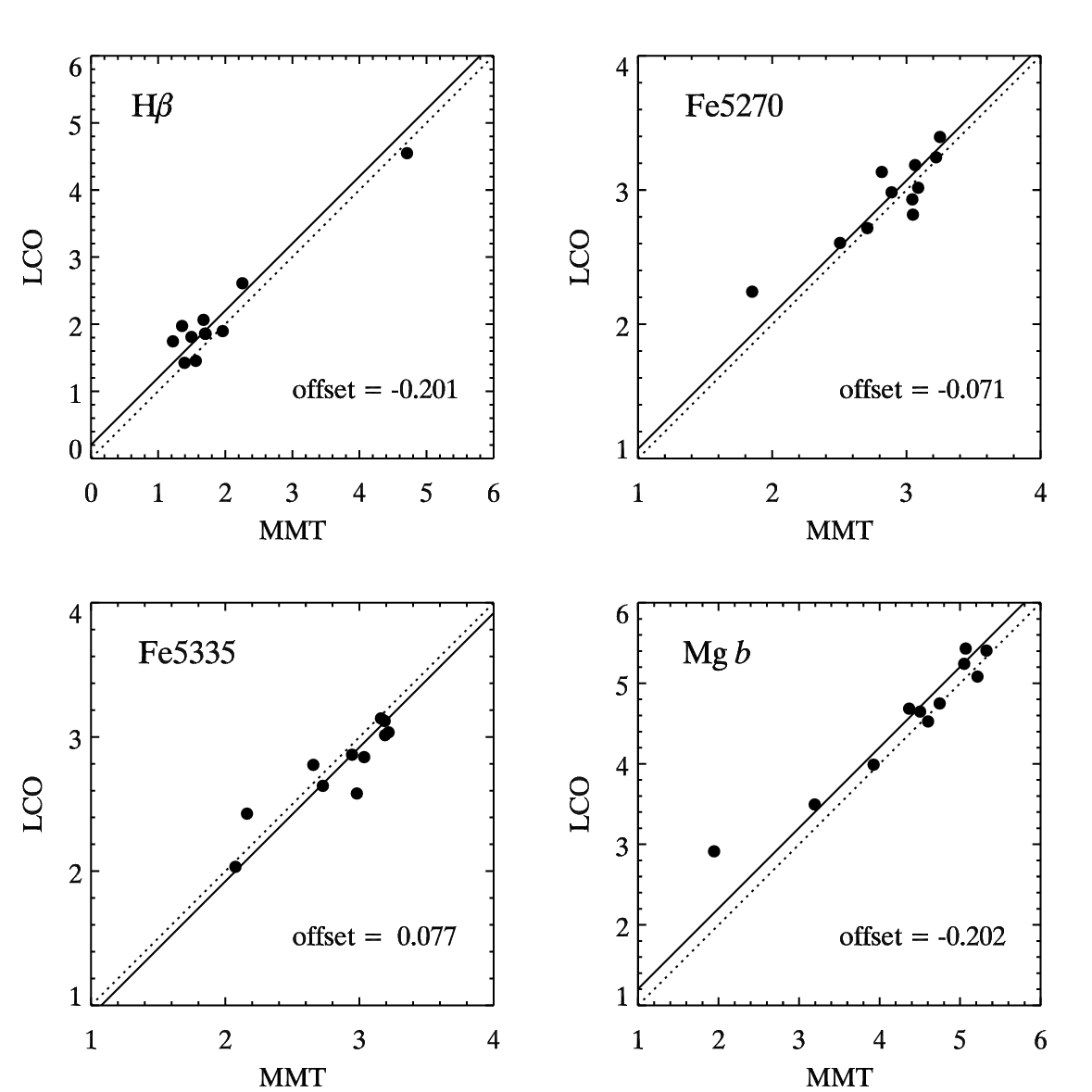}
\caption{\label{figure4}Same as Figure~\ref{figure3}, but for the common galaxies observed both at MMT ($x$-axis) and LCO ($y$-axis). The mean offset (MMT-LCO observations) is given in each panel.}
\end{figure}

\begin{figure}
\center
\epsscale{1.1}
\plotone{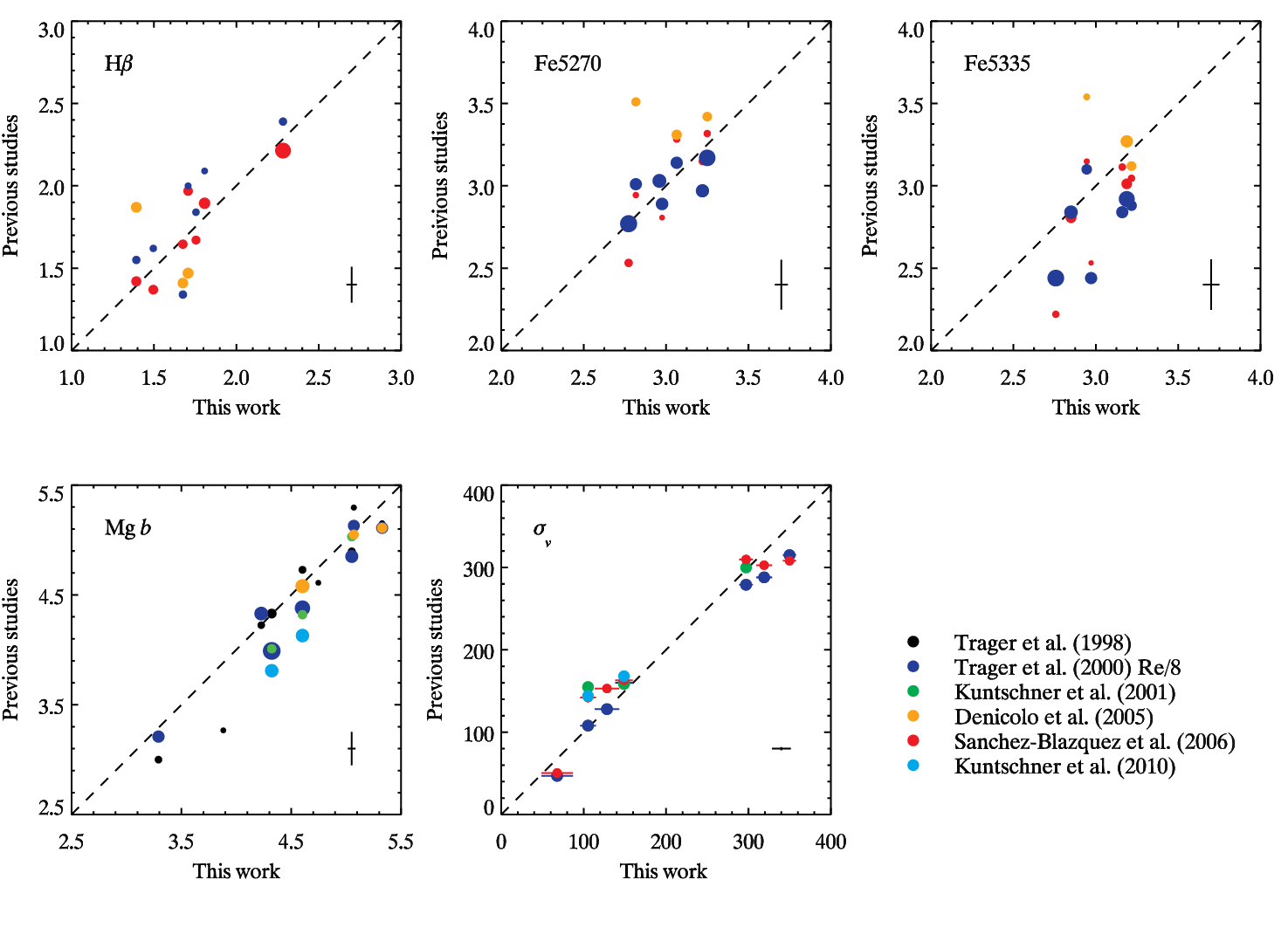}
\caption{\label{figure5}Comparison of Lick indices (H$\beta$, Fe5270, Fe5335, and Mg\,b in \AA) and velocity dispersion ($\sigma_{v}$ in km s$^{-1}$) measured in this study for the nonhost early-type galaxies observed at MMT 6.5 m with those from the literature. The dashed lines represent the one-to-one relations. The size of the symbol is inversely proportional to the error, and the typical errors for each value are indicated in the lower right corner of each panel.}
\end{figure}

For the data reduction, we followed the same procedures described in Paper I---the IRAF\footnote{IRAF is distributed by the National Optical Astronomy Observatory, which is operated by the Association of Universities for Research in Astronomy (AURA) under cooperative agreement with the National Science Foundation.}  based routines for the basic preprocessing and for the long-slit spectroscopic analysis. To obtain reliable stellar population age and metallicity, the typical S/N of our target galaxies was aimed to be unusually high, $\sim$175 per pixel (at 5000 {\AA}), and thus, even the faintest galaxies in our sample have S/N of more than $\sim$65. From our fully calibrated spectra, possible contaminations by weak emission lines, if any, were removed using the Gas AND Absorption Line Fitting \citep{Sarzi2006} package based on the Penalized Pixel-Fitting \citep{Cappellari2004} method. Most of the parameters and setups for this procedure were following the descriptions in Paper I. As an example, Figure~\ref{figure2} compares the absorption-line features for NGC~2258, before and after the emission correction. We measured the Lick indices for host galaxies using {\it lick\_ew} procedure in {\it EZ\_ages} package \citep{Graves2008} from the emission cleaned spectra, with a correction for the velocity dispersion broadening. We refer the reader to Paper I for the details of the procedures adopted in our data reduction. The measured indices were then transformed to the Lick/IDS standard system using the standard stars in \citet{Wortheyetal1994} and \citet{Worthey1997}. Figure~\ref{figure3} shows the mean offset of each index between our MMT observations and the Lick/IDS database, which was used for the zero-point correction. As the last step, we have combined the two data sets from our observations made at LCO and MMT. For this, the Lick indices of 11 galaxies common in both data sets were compared as shown in Figure~\ref{figure4}, from which the measured values of MMT spectra were shifted to the LCO system. In Figure~\ref{figure5}, our measured indices for nonhost ETGs observed at MMT are compared with those reported in the literature \citep{Trager1998,Trager2000,Kuntschner2001,Kuntschner2010,Denicolo2005,Sanchezblazquez2006}. Similarly to the result reported in Paper I, our measurements are consistent with previous measurements to within -0.03, -0.12, 0.070, 0.20 \AA, and $\sim$10 km/s for H$\beta$, Fe5270, Fe5335, Mg\,$b$, and $\sigma$$_{v}$, respectively. These small offsets are comparable to those generally shown in other studies and are expected from the differences in the adopted aperture size, instruments used, and/or the procedures employed for the emission correction. The fully corrected Lick indices (H$\beta$, Fe5270, Fe5335, and Mg\,$b$), together with the measured velocity dispersion ($\sigma$$_{v}$), for our total sample are listed in Table~\ref{table4}.

\section{Correlation between Population Age and Hubble Residual of SN Ia}

\begin{figure}
\center
\epsscale{1.1}
\plotone{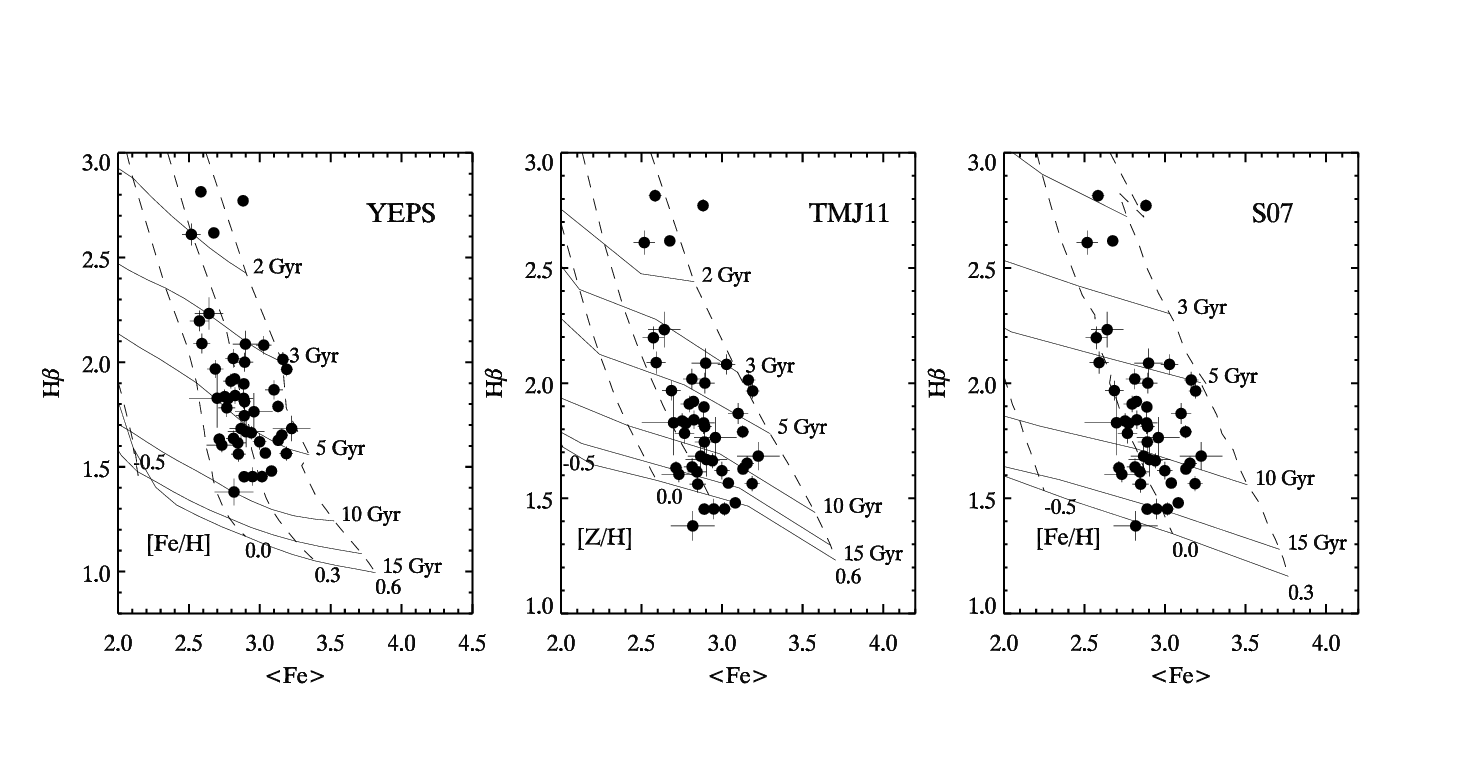}
\caption{\label{figure6}Determination of luminosity-weighted mean age and metallicity for our sample of early-type host galaxies. Filled circles represent measured indices for our host galaxy sample which are overlaid with their errors on the model grids (${\rm[\alpha/Fe]}$ = 0.3) from three different sets of EPS models (YEPS, TMJ11, and S07). The solid horizontal lines are for common ages, while the dashed vertical lines are for the same metallicity ([Fe/H] for YEPS and S07, [Z/H] for TMJ11).}
\end{figure}

\begin{figure}
\center
\epsscale{0.8}
\plotone{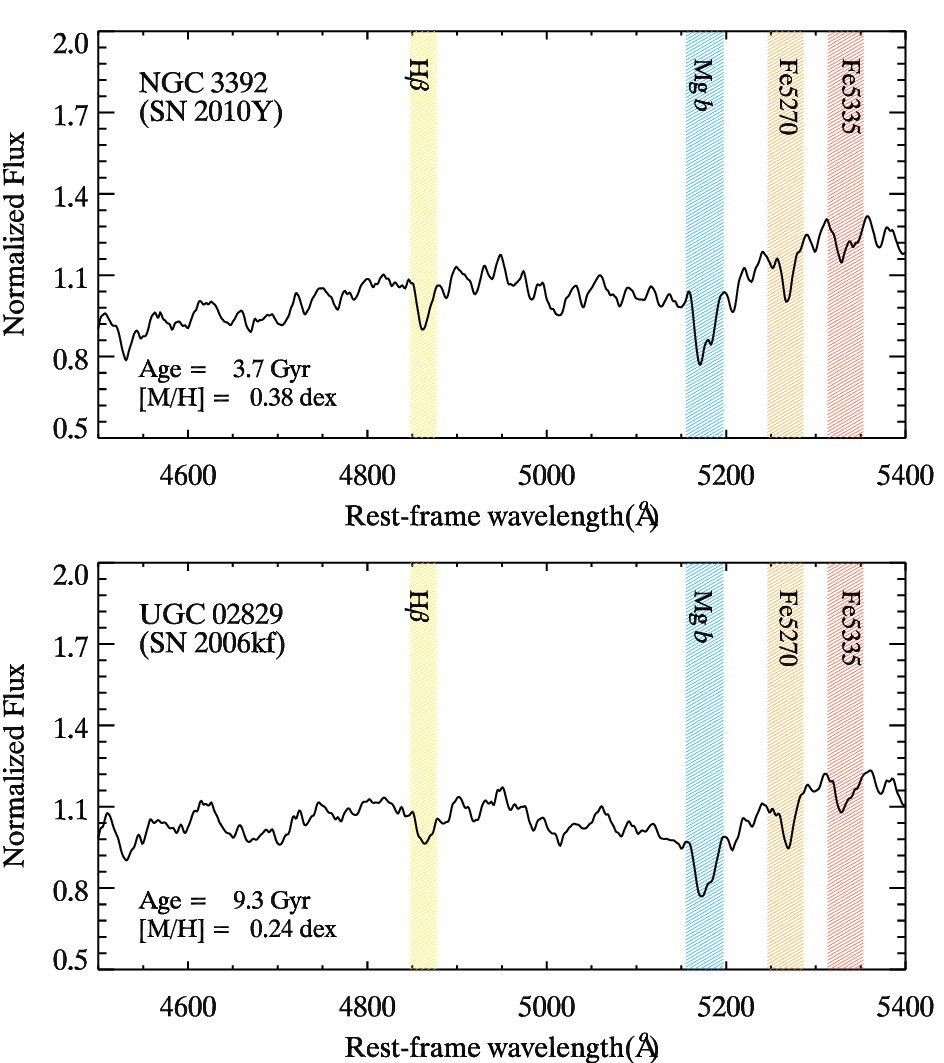}
\caption{\label{figure7}Example of typical spectra for the relatively young (NGC~3392; upper panel) and old host galaxies (UGC~02829; lower panel). Note the difference in the depth of H$\beta$ index. Population age and metallicity are given in each panel. Colored shades are absorption bands for H$\beta$, Mg\,$b$, Fe5270, and Fe5335, respectively.}
\end{figure}

With the measured Lick indices, H$\beta$, Mg\,$b$, and $\langle$Fe$\rangle$,\footnote{$\langle$Fe$\rangle$ $\equiv$ (Fe5270 + Fe5353)/2} here we determine the luminosity-weighted age and metallicity of stellar population composing the central region of host galaxies. For this, we adopt three different evolutionary population synthesis (EPS) models from \citet[][Yonsei Evolutionary Population Synthesis, hereafter YEPS]{Chung2013}, \citet[][hereafter TMJ11]{Thomas2011}, and \citet[][hereafter S07]{Schiavon2007}. As described in detail in Paper I, age and metallicity ([M/H]) are simultaneously determined from these models on the  $\langle$Fe$\rangle$-H$\beta$ model grid. Figure~\ref{figure6} compares the measured $\langle$Fe$\rangle$ and H$\beta$ indices of our host galaxy sample with each of these EPS models for ${\rm[\alpha/Fe]}$=0.3. Table~\ref{table5} lists the derived values for population age and [M/H], together with their errors, from YEPS, TMJ11, and S07 models, respectively. In Figure~\ref{figure7}, two example spectra for relatively young and old host galaxies are compared, where one can see a difference in the depth of H$\beta$ line between these two galaxies.

\begin{figure}
\center
\epsscale{0.8}
\plotone{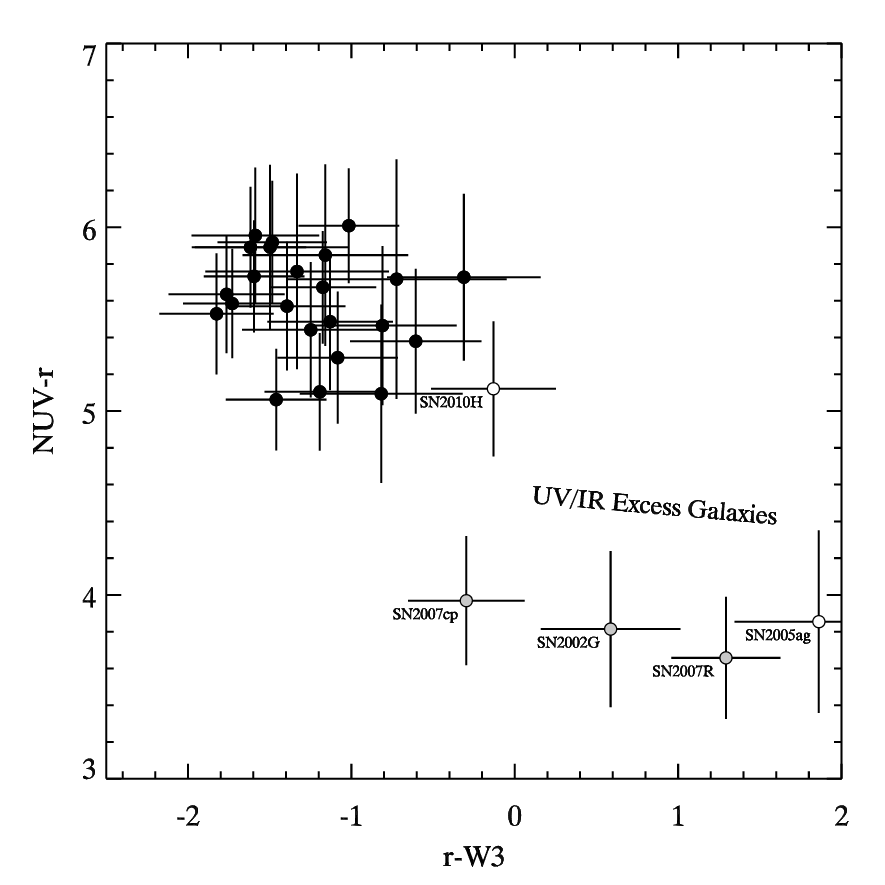}
\caption{\label{figure8}UV-optical-IR color-color diagram of the host galaxies used to select out nongenuine early-type galaxies showing features of the recent episode of starburst. The gray circles are for galaxies showing excess emissions both in $NUV$ and $W3$, and the white circles denote the rejuvenated galaxies (age $<$ 2.5 Gyr). Mean isophotal colors at the effective radii are measured from the PSF-matched \citep[][]{Aniano2011} multi-wavelength image.}
\end{figure}

For the purpose of this paper, we need to select only genuine ETGs with absorption-dominated spectra from passive environments without ongoing star formation. However, some galaxies in our sample, although they are morphologically classified as ETGs, show strong emission lines on their spectra, indicating they have more characteristics of late-type environments than early-type systems. They appear as if they have very young ages ($<$2.5 Gyr) after strong emission line correction. According to \citet{Thomas2010}, such galaxies apparently younger than 2.5 Gyr in terms of luminosity-weighted age have most likely been ``rejuvenated'' by recent episode of star formation \citep[see also][]{Yi2005,Shapiro2010}. For these rejuvenated galaxies, as demonstrated by many investigators \citep[see, e.g.,][]{Greggio1997,Nolan2007,Serra2007}, only a few percent of very young stellar population can significantly affect the luminosity-weighted age. Therefore, the majority ($>$90\%) of stellar populations in these galaxies can still be markedly older than the determined mean age. In our sample, six galaxies, including NGC~5018 \citep{Nolan2007, Spavone2018}, are classified as such rejuvenated galaxies. They also show star-forming rings, gas lanes, and/or disturbed features on their DSS image (see Figure~\ref{figure1}), all indicating features of recent episodes of merger-induced starburst. Additional IFU observations and analyses would be required for these galaxies to dissect into young and old components, which, however, is still a very challenging task \citep[see, e.g.,][]{Guerou2016,Zibetti2019}. For this reason, here we have excluded these rejuvenated galaxies from further analyses because their mean ages are seriously underestimated or highly uncertain. In addition to these galaxies, some of our host galaxies can still contain a small fraction of young stellar component. To check this, when available, we have examined the level of young star contamination in our host sample by using the $NUV$-, $r$-, and $W3$-band images from $GALEX$ \citep{Martin2005}, SDSS \citep{Abolfathi2018} or DESI Legacy Imaging Surveys \citep{Dey2019}, and unWISE \citep{Lang2014} databases. We find that three galaxies (hosts of SN~2002G, SN~2007cp, and~SN 2007R) have very large excess emissions both in $NUV$ and $W3$, clearly separated from other host galaxies (see Figure~\ref{figure8}), indicating a significant contamination from ongoing or recent star formation. Along with the rejuvenated galaxies, these UV/IR excess galaxies are further excluded from our final analyses as their derived ages are highly uncertain.

Following our Paper I, in Figure~\ref{figure9}, we first compare the velocity dispersion (a proxy of galaxy mass) with population age and metallicity for our final sample of host galaxies. As in our previous analysis, we perform the linear fitting using the Markov Chain Monte Carlo (MCMC) analysis implemented in the LINMIX package \citep{Kelly2007}. Host mass and population age show a strong correlation, which corresponds to $>$99.9\% probability for the non-zero slope in all cases of adopted EPS model \footnote{To estimate the confidence level, we fit the posterior distribution of slopes with a Gaussian function, and computed the standard deviation of the slopes and computed how many sigma the mean is away from zero, then converted that to a probability.}. This is consistent with the result reported in Paper I, but the present result is based on a lager sample. In contrast, the metallicity shows no meaningful correlation with velocity dispersion (see Figure~\ref{figure10}).

It is well-established that SN light-curve parameters (decline rate or stretch factor) are correlated with peak luminosity before standardization \citep{Phillips1993,Riess1999,Hamuy2000,Howell2001,Sullivan2006}. Previous studies based on indirect or secondary age indicators (spectral energy distribution (SED), SFR, morphology) for host galaxies \citep[e.g.,][]{Howell2009,Neill2009,Johansson2013,Rose2019} also found a strong correlation between the stretch factor and population age, in the sense that younger population tends to host intrinsically brighter SN Ia. In Figure~\ref{figure11}, we plot SALT2 stretch factor ($X_{1}$) with population age for our host galaxy sample, which shows qualitatively consistent correlation with previous studies, although the statistical significance is lower ($\sim$86\% confidence level). This is probably because our result is based only on ETGs with relatively narrow spans of age (2.5 Gyr---10.5 Gyr with YEPS) and the light-curve stretch value.

As described above, the SN brightness after light-curve shape and color corrections, instead of the intrinsic luminosity, is being used in SN cosmology, and therefore, it is crucial to see the correlation between stellar population age and standardized SN brightness. Figure~\ref{figure12} shows a correlation of stellar population age with the HR calculated from standardized SN Ia brightness. Note that, on this diagram, a positive value of HR indicates a relatively fainter SN Ia after the standardization, while a negative value corresponds to a brighter SN Ia. Interestingly, at first glance, no sample galaxy appears to be located in the upper-right regime of the diagram, which would imply a correlation between age and standardized SN brightness. \citet{Rose2019} also noted a similar distribution from their study with SED-based age dating. In Figure~\ref{figure13}, we perform the MCMC analysis with our final sample (excluding rejuvenated and UV/IR excess galaxies). We find that the population age is correlated with the standardized brightness (HR) at a $\sim$99.5\% confidence level. Note that our result is insensitive to the choice of population synthesis model, as they all show qualitatively consistent results in the sense that SN Ia in younger host tends to be fainter than that in older host. It is important to note, from the comparison of Figure~\ref{figure13} with Figure~\ref{figure12}, that contamination of nongenuine ETGs(rejuvenated and UV/IR excess galaxies) would make the correlation apparently weaker. Therefore, careful selection of only normal ETG sample is important for this study. In that respect, it is reassuring to see that the correlation is getting stronger when the contamination from nongenuine ETGs is properly taken into account. For the quantitative analysis in the following section, we will adopt the result from YEPS because it is based on the more realistic treatment of Helium burning stage in the stellar evolution modeling and is well-calibrated to the color-magnitude diagrams, integrated colors, and absorption indices of globular clusters in the Milky Way and nearby galaxies \citep{Lee2005,Chung2013,Chung2017,Joo2013,Kim2013}. It is also producing ages for oldest host galaxies which are not inconsistent with the age of the universe. Unlike population age, we find no significant correlation between host metallicity and HR (see Figure~\ref{figure14}). In Table~\ref{table6}, we summarize the results of our statistical analyses.

\section{Discussion}\label{Section4}
In this paper, we have made a critical test on the possibility of the luminosity evolution in SN cosmology by investigating the correlation between SN Ia brightness and stellar population age of early-type host galaxies. Based on the most reliable age dating from high-quality absorption-dominated spectra and well-established stellar population models, we found a significant correlation between standardized SN brightness and stellar population age at $\sim$99.5\% confidence level. As such, this is the most direct and stringent test ever made for the luminosity evolution of SN Ia. This result makes it possible to shed new light on the origin of the previously reported correlations of SN Ia luminosity with host morphology, host mass, and local SFR. Below we discuss how our result is related and could be accommodated with these previous studies.

As briefly described in Section~\ref{Section1}, the very first test for the possible luminosity evolution of SN Ia, after the light-curve corrections, was to use the host galaxy morphology as a proxy for stellar population age \citep{Riess1998,Schmidt1998,Perlmutter1999}. Although this possibility was initially ruled out from no apparent dependence of SN luminosity with a limited sample of host morphologies, more recent analysis with a larger sample shows a different result \citep{Hicken2009}. This latter study, where the effect of dust extinction was carefully corrected, reported that SNe occurred within the late-type (Scd-Irr) galaxies are $\sim$0.14 mag fainter than those hosted by the early-type (E-S0) galaxies (see also \citealt{Suzuki2012} for a similar result). If their result is interpreted as originally intended in the discovery papers, this dependence must indicate the difference in stellar population age. That the stellar populations in late-type galaxies are younger, in the mean, than those in ETGs is a well-established result in extragalactic studies. According to the recent studies based on high-quality spectroscopic data \citep[see, e.g., Figure 6 of][]{Scott2017}, stellar populations in late-type (Scd/Sd) galaxies are in the mean $\sim$4 Gyr younger than those in early-type (E/S0) galaxies. The most plausible interpretation of the result of \citet{Hicken2009} is therefore that a SN Ia occurred at a younger host galaxy is fainter than that found in an older host galaxy ($\sim$0.14 mag/4 Gyr). It is interesting to see that this conversion to age effect is not only consistent with but also quantitatively comparable to our result from population age and SN Ia luminosity (see Figure~\ref{figure13} and Table~\ref{table6}). 

\begin{figure}
\center
\epsscale{1.1}
\plotone{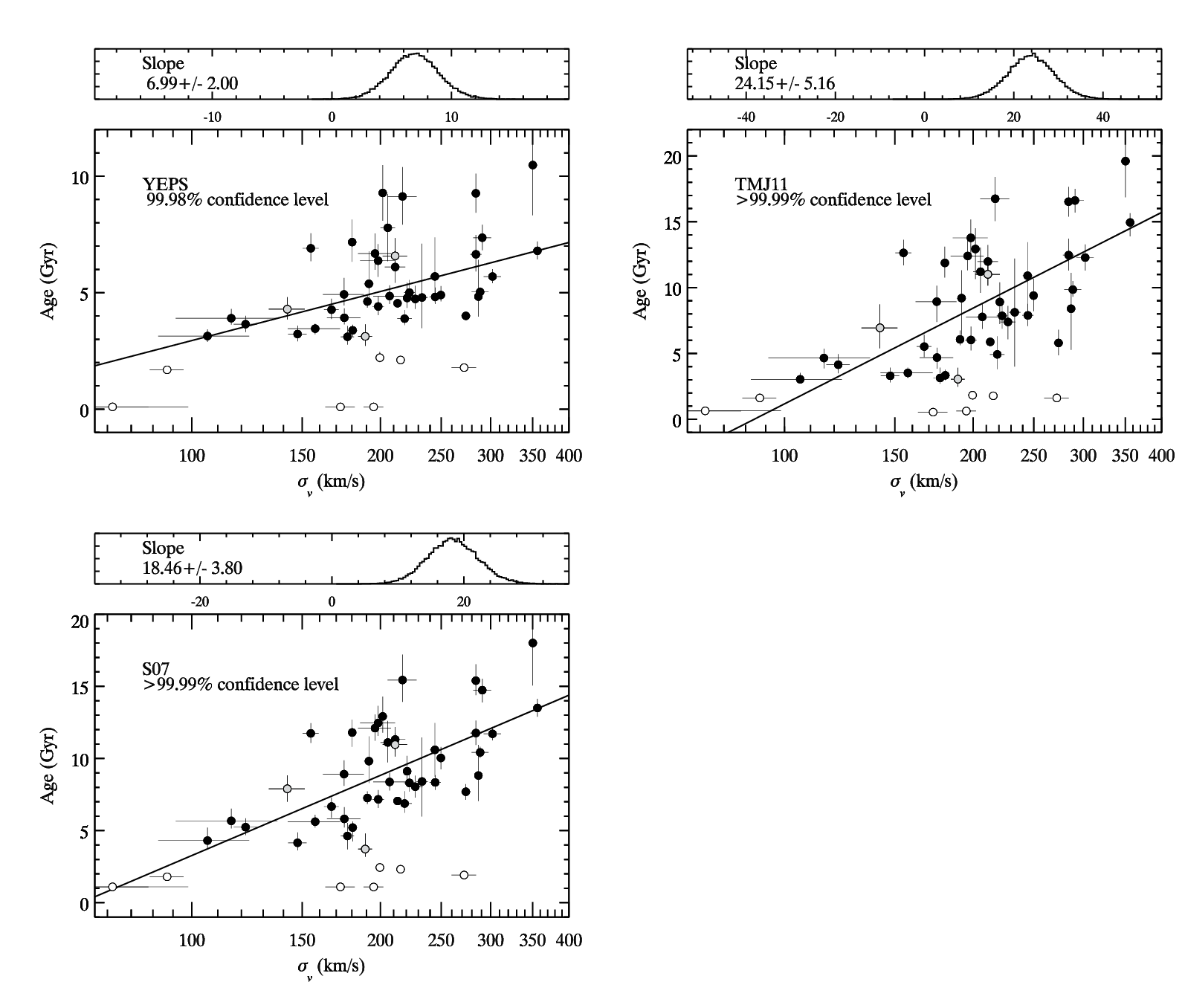}
\caption{\label{figure9}Correlation between velocity dispersion ($\sigma$$_{v}$) and stellar population age for our sample of host galaxies. Population ages of galaxies were determined using YEPS, TMJ11, and S07 models, respectively. The black solid line is the regression line obtained from the posterior median estimate of MCMC analyses, and the confidence level of the correlation is given in each panel. Nongenuine early-type galaxies excluded from our final analyses are denoted by the white and gray circles for the rejuvenated and UV/IR excess galaxies, respectively.}
\end{figure}

\begin{figure}
\center
\epsscale{1.1}
\plotone{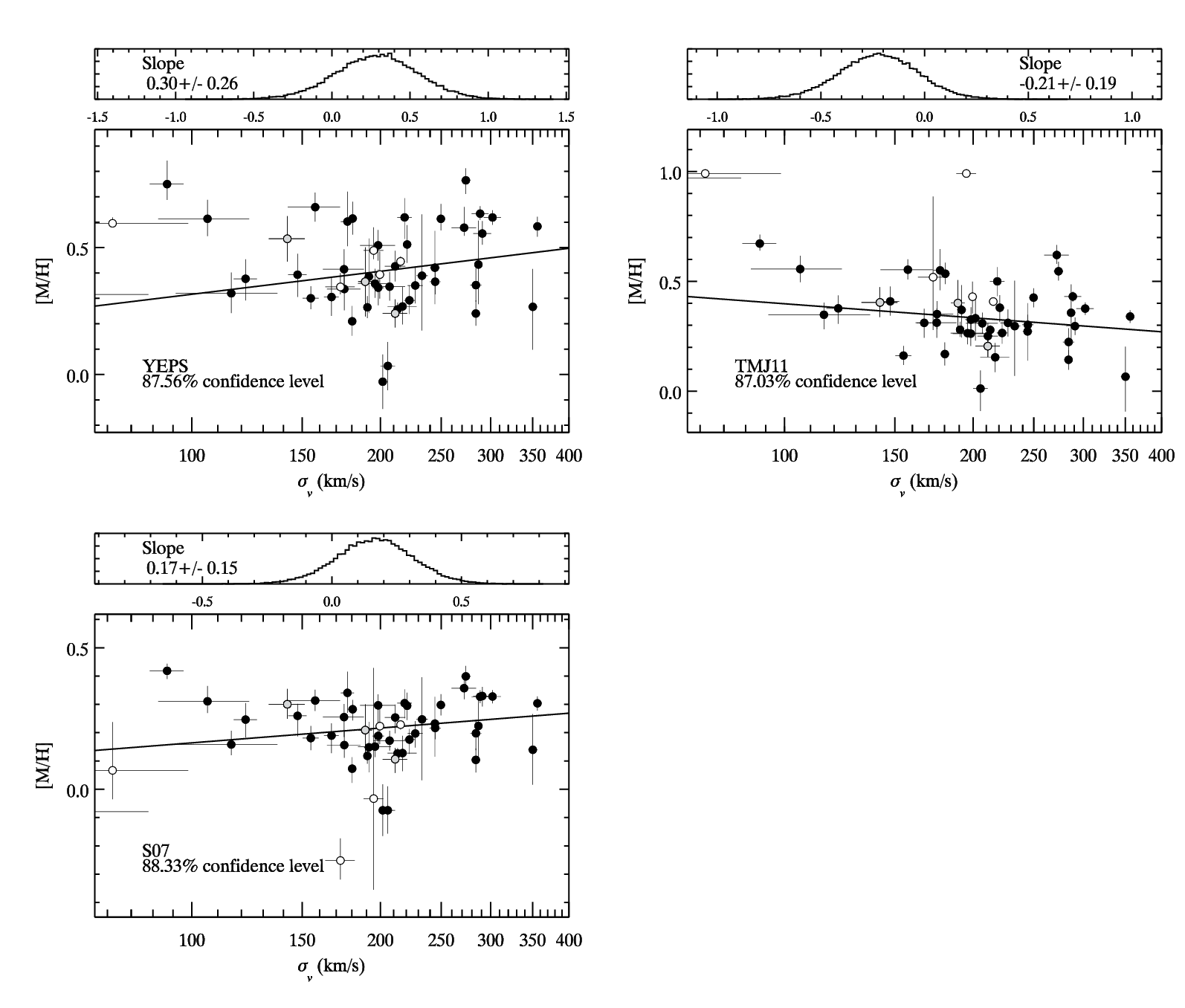}
\caption{\label{figure10}Same as Figure~\ref{figure9}, but for the metallicity [M/H] which shows no significant correlation.
}
\end{figure}

\begin{figure}
\center
\epsscale{1.1}
\plotone{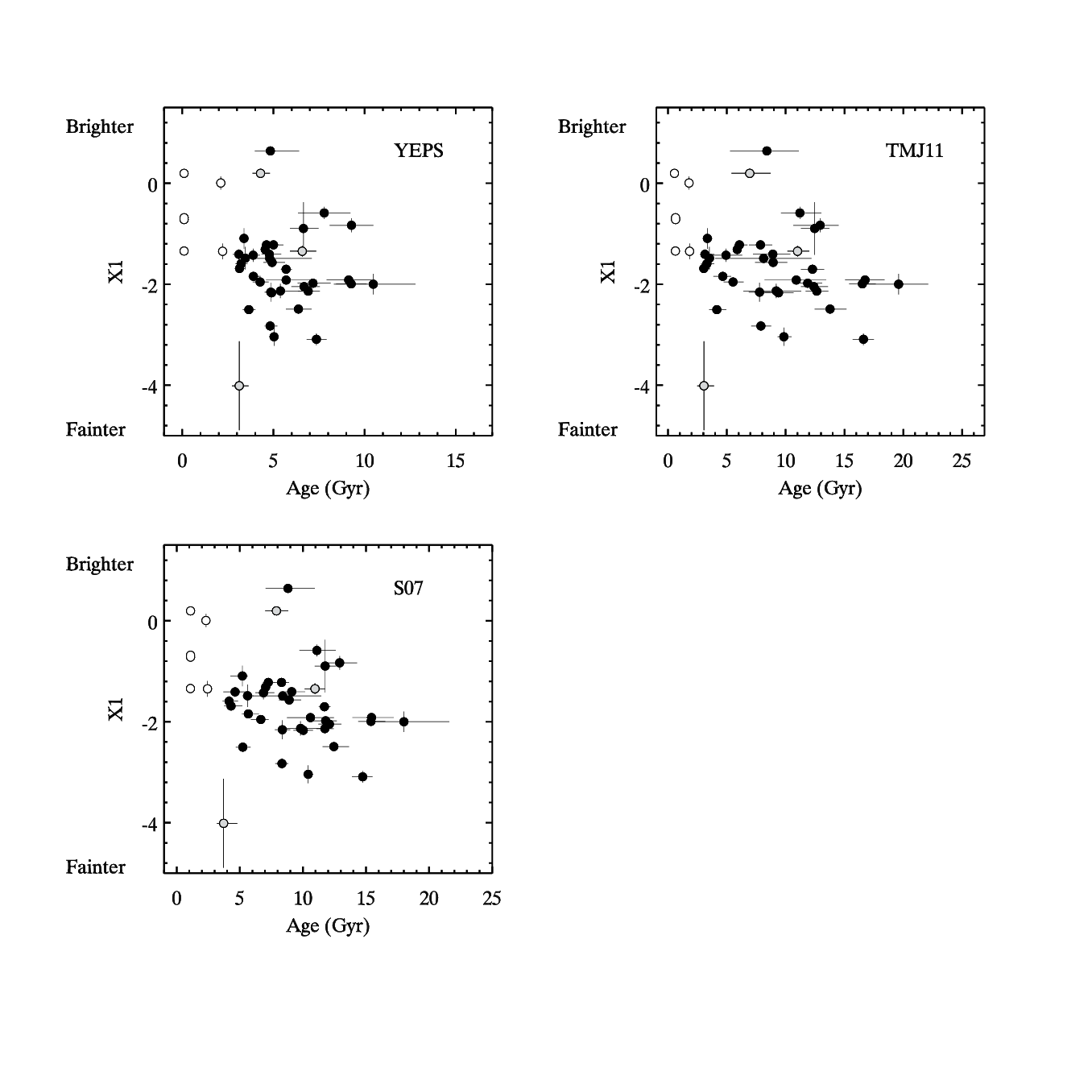}
\caption{\label{figure11}Correlation of population age for host galaxies with SN stretch factor ($X_{1}$). Population ages from YEPS, TMJ11, and S07 models are shown on each panel, respectively. Nongenuine early-type galaxies excluded from our final analyses are denoted by the white and gray circles for rejuvenated galaxies and UV/IR excess galaxies, respectively.}
\end{figure}

\begin{figure}
\center
\epsscale{1.1}
\plotone{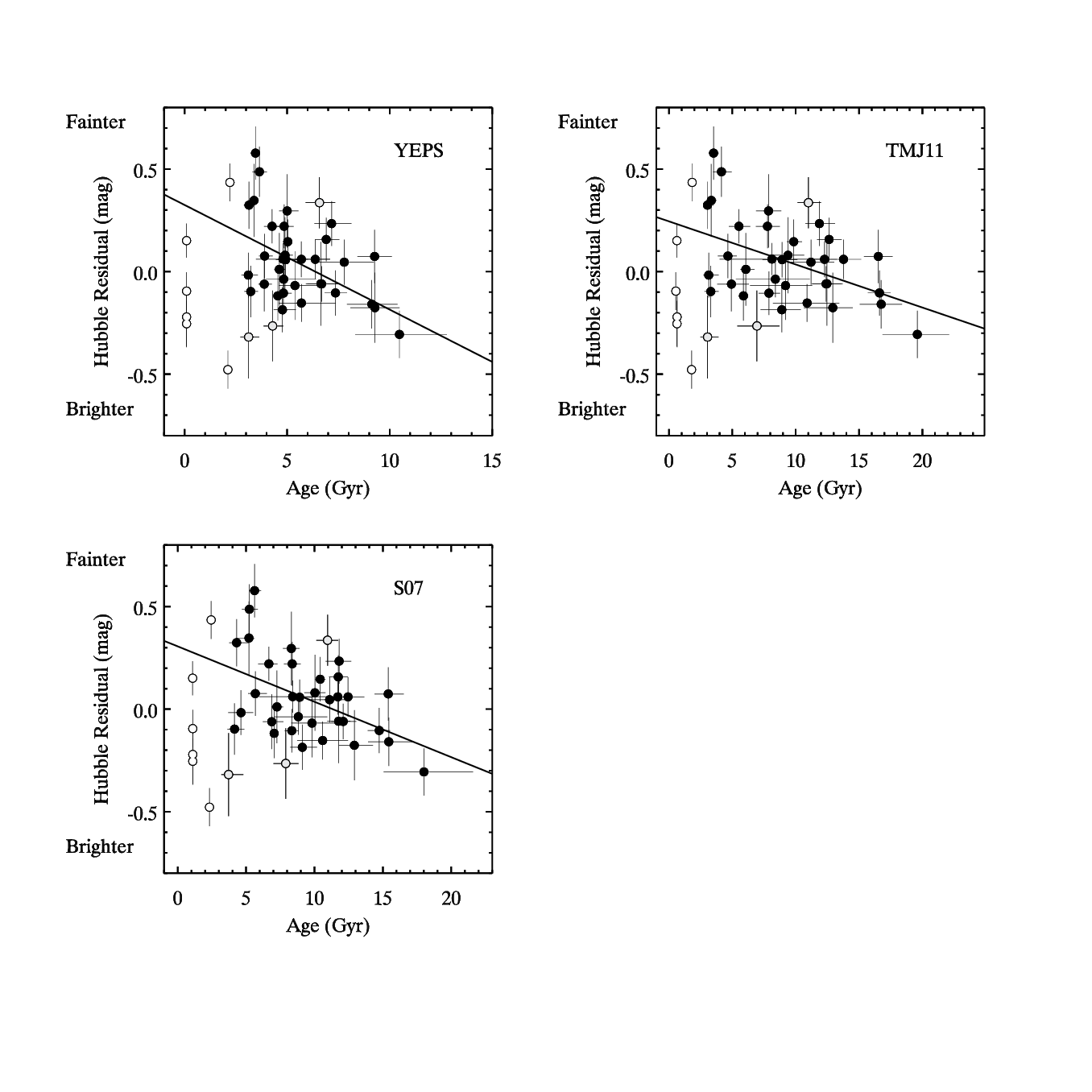}
\caption{\label{figure12}Correlation between population age of host galaxy and Hubble residual of SN Ia. Population ages from YEPS, TMJ11, and S07 models are shown on each panel, respectively. Non-genuine early-type galaxies excluded from our final analyses are denoted by the white and grey circles for rejuvenated galaxies and UV/IR excess galaxies, respectively. The solid line is the regression fit obtained from MCMC analysis only for our final sample.} 
\end{figure}

\begin{figure}
\center
\epsscale{1.1}
\plotone{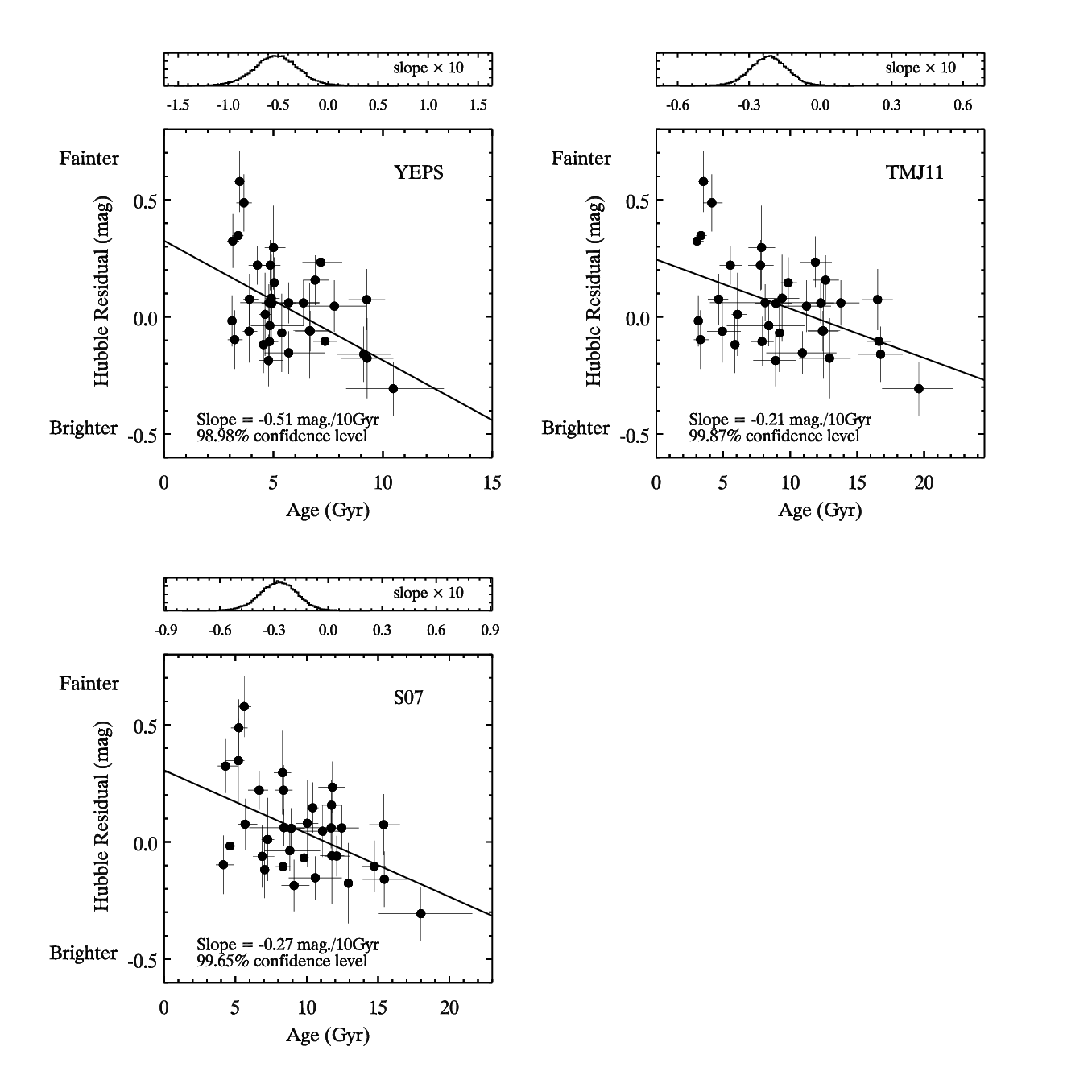}
\caption{\label{figure13}Same as Figure~\ref{figure12}, but only for our final sample. The difference in Hubble residual per 10 Gyr and the confidence level are given in each panel together with the posterior slope distribution.}
\end{figure}

\begin{figure}
\center
\epsscale{1.1}
\plotone{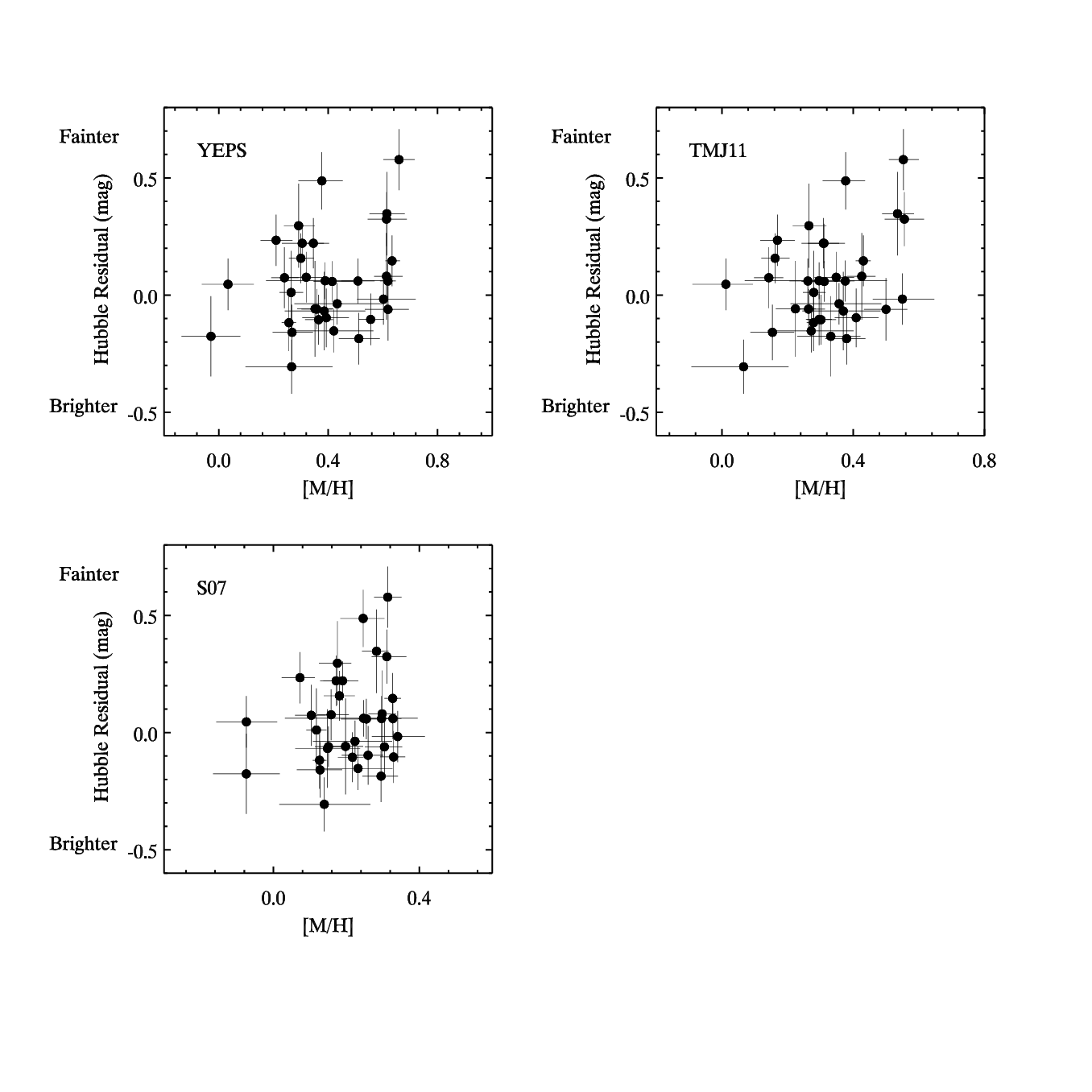}
\caption{\label{figure14}Same as Figure~\ref{figure13}, but for the metallicity [M/H] which shows no significant correlation.}
\end{figure}

The mass of a host galaxy was then adopted as a proxy for population properties in many host galaxy studies. Among these, careful investigations by \citet{Sullivan2010} and \citet{Kelly2010} have established that SNe Ia in less massive galaxies (by a factor of 10) are $\sim$0.08 mag fainter than those in more massive galaxies. As the mass of a galaxy cannot directly affect the brightness of SN Ia occurred in a host, while the mass of a progenitor would have a direct impact on the intrinsic luminosity of SN Ia \citep{Umeda1999,Howell2009,Johansson2013}, this correlation must be originated from either population age or metallicity. Among nonhost galaxies, there is a well-established correlation between galaxy mass and stellar population age \citep{Thomas2005, Neistein2006}. We have shown in Paper I that host galaxies are also following this ``downsizing'' trend with no apparent dependence on metallicity. The same result based on a larger sample in Figure~\ref{figure9} shows that less massive host galaxies (by a factor of 10) are $\sim$2 Gyr younger (based on YEPS) than more massive galaxies. Therefore, the apparent correlation with host mass is, again, most likely due to the effect of stellar population age ($\sim$0.08 mag/2 Gyr), which is also consistent with the results from our direct age dating and from the morphological classification described above.

More recent studies based on spectroscopy measured a local SFR in the vicinity of the site where SN Ia arose in a host galaxy. The local SFR can be readily and reliably obtained from the emission lines which are observed in a significant fraction of host galaxies \citep{Rigault2013,Rigault2015,Rigault2018,Galbany2014}. Among these studies, \citet{Rigault2018} showed that SNe Ia in locally star-forming environments (higher local SFR) are $\sim$0.16 mag fainter than those in locally passive environments (lower local SFR), after the conventional light-curve corrections. As the physics involved in SFR would not change the SN luminosity directly, this correlation is also most likely due to the stellar population age. \citet{Rigault2018} also suggested that this difference is originated from the different fraction of young to old stars between the star-forming and the passive environments. In this respect, more detailed analysis for the typical SN progenitor ages can be performed for the two environments. For the locally star-forming environment, first, we adopt the stellar population age from \citet{Galbany2014} which was derived from the spectral fitting technique with the spectral synthesis code STARLIGHT \citep{CidFernades2005}. Out of 36 host galaxies with normal SN Ia in their sample, 10 galaxies belong to the locally star-forming group ($\mbox{$\Sigma$}SFR_{\text{SN}}$ $>$ 10$^{-2}$ or log~$M_{\star}$ $<$ 10) as defined by \citet{Rigault2018}, and for them we obtain the mean age of $\sim$0.6 Gyr. A similar quantity is also derived by convolving the SN Ia delay time distribution (DTD) with a local star formation history (SFH) as demonstrated by \citet{Childress2014}. For this, we employ a SFH of the Milky Way disk (the solar neighborhood) from the analyses of \citet{Bernard2018} as a representative for the star-forming environment. When this SFH is coupled with a DTD of SN Ia suggested by \citet{Childress2014}, the peak of the SN progenitor age distribution is also found at $\sim$0.5 Gyr. As to the locally passive environment ($\mbox{$\Sigma$}SFR_{\text{SN}}$ $<$ $10^{-2}$ and log~$M_{\star}$ $>$ 10), 19 galaxies in \citet{Galbany2014} belong to these criteria, and for this group, we obtain the mean age of $\sim$2.3 Gyr. Six galaxies in our sample (Table~\ref{table1}) are in common with the list of host galaxies in \citet{Childress2013}, and for these ETGs (locally passive environments), the mean population age from our study is $\sim$3.8 Gyr.  By taking the mean value of the two analyses for each environment, we adopt $\sim$0.6 Gyr as a typical age of the stellar population in locally star-forming environments and $\sim$3.1 Gyr for that in locally passive environments. Therefore, the 0.16 mag difference between the locally star-forming and locally passive environments is most likely due to a $\sim$2.5 Gyr difference in stellar population age. This difference is still comparable with the result we obtained from our early-type host galaxies.

Table~\ref{table7} summarizes the results from these analyses compared with the slope ($\Delta$HR/$\Delta$age) derived from our direct age dating. 
The last column lists the converted value from each of the host properties to the age difference. Note that, when interpreted as a population age effect, these four different and independent approaches are all pointing to the same direction --- SNe Ia in younger environments are fainter than those in older environments, after the standardization of SN brightness. It is important to note that all of the three well-established correlations between the SN brightness and host property are not only qualitatively consistent but also quantitatively comparable to our result based on population age. Furthermore, all of these host properties (morphology, mass, and local SFR) are closely interrelated as described above. Based on these, we believe that the correlation with population age is most likely the original correlation that drove the previously reported correlations with host properties. This appears to be the only plausible interpretation that can explain all of these correlations (including our own) simultaneously.

\begin{figure}
\center
\epsscale{0.8}
\plotone{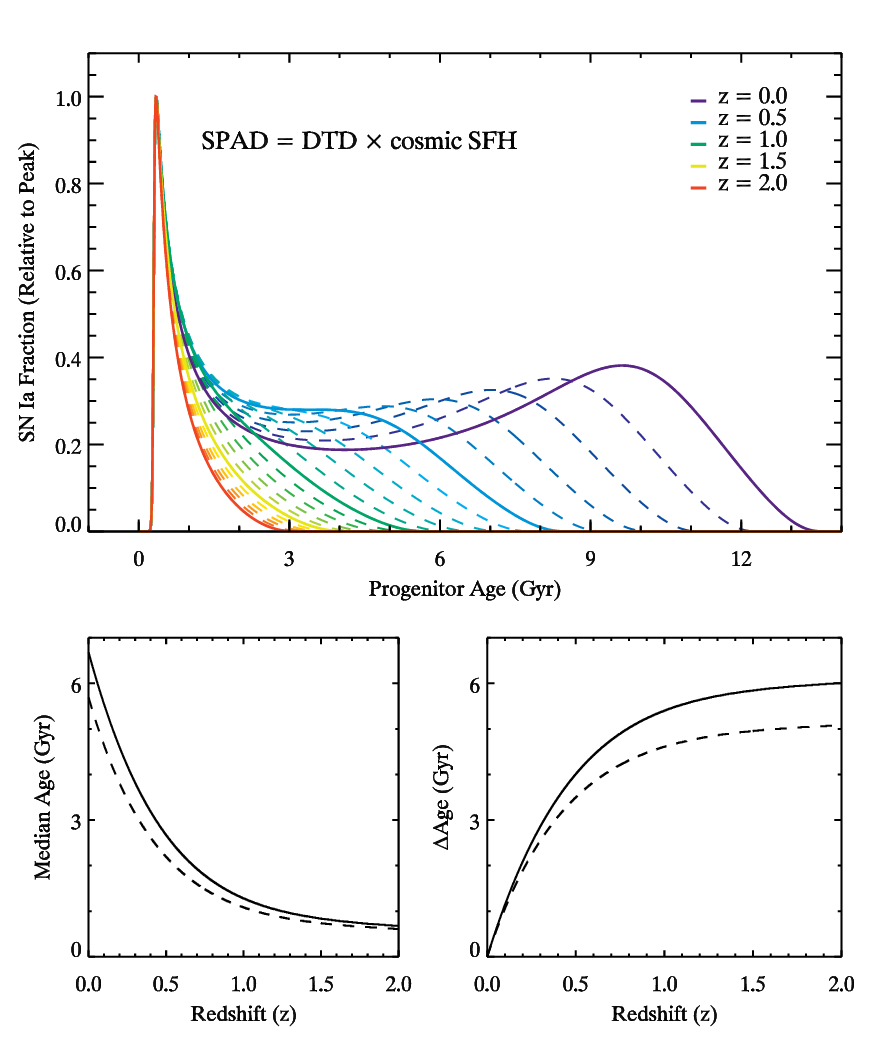}
\caption{ \label{figure15} SN Ia progenitor age distribution (SPAD) as a function of redshift calculated following \citet{Childress2014}. This is for the sample of SNe Ia arising at a given epoch of cosmic time summed over all galaxy types. The upper panel shows the distributions in steps of $\Delta z$ = 0.1. The lower panels are the median value of the age distribution (left) and the difference with respect to $z$ = 0 (right). To calculate a look-back time at a given redshift, we implicitly adopted the $\Lambda$CDM cosmological model ($\Omega_{M}$ = 0.27, $\Omega_{\Lambda}$ = 0.73), while the dashed lines in the lower panels show the results for the cosmological model without $\Lambda$ ($\Omega_{M}$ = 0.27, $\Omega_{\Lambda}$ = 0.00). }
\end{figure}

Our result, together with the age interpretation of the previously reported correlations, inevitably suggests the SN luminosity evolution with redshift. This is because, as is well-known, host galaxies at $z$ $\sim$ 0.0 consist of both young and old stellar populations whereas those at high redshift only consist of young population.  Previous studies based on host mass and local SFR only considered the redshift evolution of these quantities, instead of population age, and therefore the applied corrections on the SN luminosity with redshift were negligible. To estimate the effect of luminosity evolution based on our result, we first derive the SN Ia progenitor age distribution as a function of redshift following the methodology in \citet{Childress2014}. The SN Ia progenitor age distribution at epoch $t_{0}$ of a given redshift $z$ of the SN exploding from a progenitor system of age $\tau$, $P(\tau; t_{0})$, is derived by convolving DTD $\phi$($\tau$) of SN Ia with cosmic SFH $\psi(t_{0} -\tau)$ in the form of $P(\tau; t_{0}) = \phi(\tau)\psi(t_{0} -\tau)$ \citep[see Figure 3 of][]{Childress2014}. Specifically, we use a smooth function for DTD suggested by \citet{Childress2014}:\\
\begin{equation}
\phi(t) \propto {{(t/t_{p})^{\alpha}} \over {{(t/t_{p})}^{{\alpha} -s} +1}} ,
\end{equation}\\
where the prompt timescale ($t_{p}$) is adopted to be 0.3 Gyr and the power-law slope ($s$) is set to -1 with $\alpha = 20$. As for the cosmic SFH, we also adopt an empirical function derived by \citet{Behroozi2013} from the collected cosmic SFR (CSFR) data:\\
\begin{equation}
CSFR(z) = {{C} \over {10^{A(z-z_{0})}+10^{B(z-z_{0})}}} ,
\end{equation}\\
where constants $A = -0.997, B = 0.241, C = 0.180$, and $z_{0}$ = 1.243 are as given in Table 6 of their paper. The calculated SN Ia progenitor age distribution is shown in Figure~\ref{figure15}, following the color-coding of \citet{Childress2014}. Note that the SN progenitor age distribution at $z = 0.0$ is strongly bimodal with two peaks at $\sim$9.4 Gyr and $\sim$0.4 Gyr, which would correspond to the ``delayed'' and ``prompt'' components of SNe Ia \citep{Sullivan2006,Childress2014}, whereas the distribution at $z > 0.5$ has only a single peak which converges to the peak value of the DTD at $\sim$0.4 Gyr. As discussed below, this is important stellar astrophysics that must be considered in detail in SN cosmology. In the lower panels, the median age of the distribution and the $\Delta$age with respect to $z = 0.0$ are also plotted as a function of redshift. According to this, the median ages of SN progenitors are $\sim$6.54 Gyr at $z$ = 0.0 and $\sim$1.25 Gyr at $z$ = 1.0, so the difference in population age between these two epochs is $\sim$5.3 Gyr. It is important to note that the $\Delta$age at high redshift is converging to $\sim$5.3 Gyr because the median value of the SN Ia progenitor age distribution is asymptotically approaching to the peak value of  $\sim$0.4 Gyr beyond $z$ $\sim$ 0.7.  As discussed in \citet{Childress2014}, the $\Delta$age is not strongly affected by the choice of  $t_{p}$ and ${s}$ for the SN Ia DTD.

\begin{figure}
\center
\epsscale{1.1}
\plotone{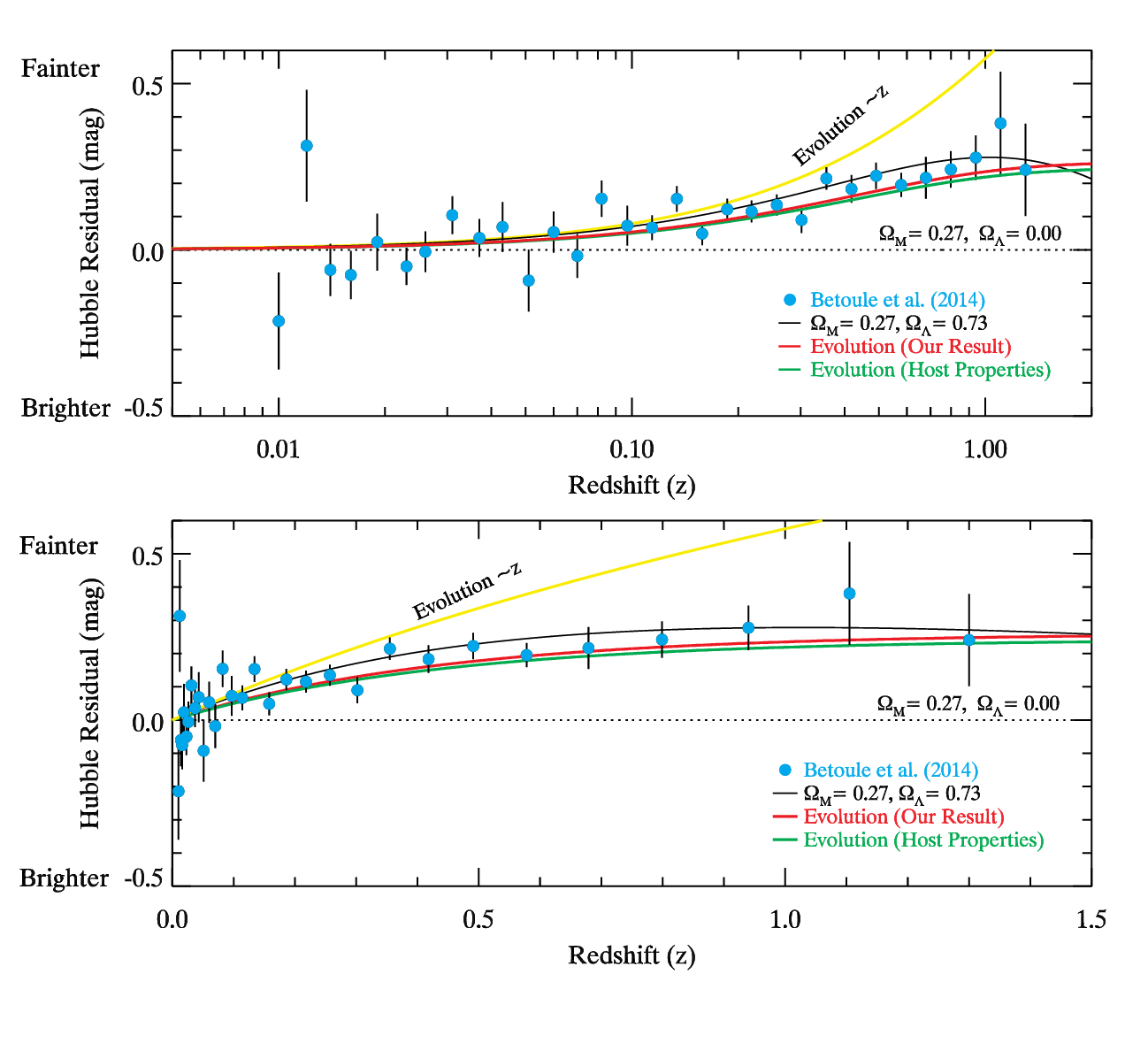}
\caption{\label{figure16} Luminosity evolution in SN cosmology predicted by our age dating of early-type host galaxies. The Hubble residuals are calculated with respect to the cosmological model without $\Lambda$ ($\Omega_{M}$ = 0.27, $\Omega_{\Lambda}$ = 0.00; the black dotted line). Assuming this model, the red line is the evolution curve based only on the age dating of early-type host galaxies, while the green line is produced using the mean value of $\Delta$HR/$\Delta$age from the four different studies on host properties. Note that our evolution curve is substantially different from that simply proportional to redshift (yellow line). The cyan circles are the binned SN data from \citet{Betoule2014}. The comparison of our evolution curves with SN data shows that the luminosity evolution can mimic a significant fraction of the Hubble residual used in the discovery and inference of the dark energy (see the black solid line).} 
\end{figure}

 In the last column of Table~\ref{table7}, we list an estimated amount of the luminosity evolution that corresponds to this age difference for each of the four different studies. The mean value of these quantities is $\sim$0.25 mag/5.3 Gyr. Over the same redshift range, the observed dimming of SN brightness in the Hubble diagram is roughly comparable to this value {\bibpunct[, ]{(}{)}{;}{a}{}{,}\citep[see, e.g.,][]{Riess1998, Hicken2009}}. Figure~\ref{figure16} shows our prediction of the SN Ia luminosity evolution in the residual Hubble diagram. The red line shows the evolution curve based only on the age dating of early-type host galaxies, while the green line is produced using the mean value of $\Delta$HR/$\Delta$age from the four different studies on host properties. As the purpose of this paper is to quantify the SN luminosity evolution with redshift and to illustrate the level of its significance, following \bibpunct[; ]{(}{)}{,}{a}{}{,}\citet[][their Figure 1]{Hicken2009} and \citet{Riess1998}, the HRs in this diagram are calculated with respect to the cosmological model without dark energy ($\Omega_{M}$ = 0.27, $\Omega_{\Lambda}$ = 0.00). The SN data over-plotted are adopted from the binned values of \citet[][their Table F.1]{Betoule2014}. Importantly, our evolution curve is substantially different from that simply proportional to redshift (see the yellow curve in Figure~\ref{figure16}) as employed by {\bibpunct[; ]{(}{)}{,}{a}{}{,} \citet[][see also \citealt{Drell2000}]{Riess2004}}. This is because the luminosity evolution of SN Ia, which is dictated by stellar astrophysics, is not linearly proportional to redshift, but determined by the SN Ia progenitor age distribution whose median value converges to $\sim$0.4 Gyr beyond $z$ $\sim$ 0.7.  Therefore, the shape of our evolution curve appears almost identical to that in redshift versus $\Delta$age diagram in Figure~\ref{figure15}.

\begin{figure}
\center
\epsscale{1.1}
\plotone{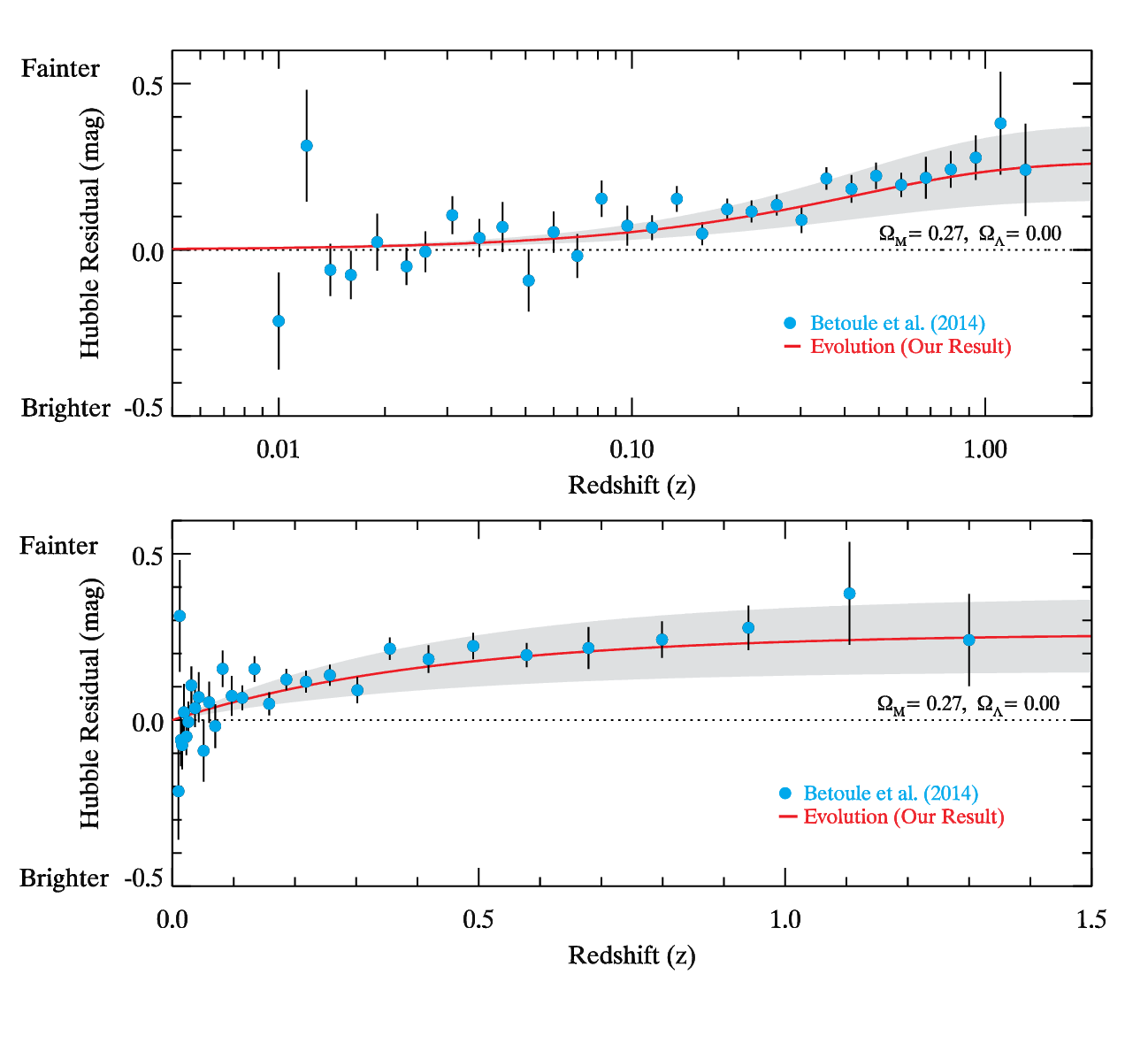}
\caption{\label{figure17}Same as Figure~\ref{figure16}, but 1$\sigma$ significance interval is added to our evolution curve (the gray shading).}
\end{figure}

Note that we implicitly assume a linear correlation between population age and Hubble residual over the full redshift range in Figure~\ref{figure16} for all SNe Ia used in cosmological sample. This assumption could be somewhat uncertain at high redshift, because our SN Ia sample has been restricted predominantly to those with low stretch factor ($X_{1} \lesssim 0.0$) preferentially discovered in early-type galaxies. While a substantial fraction (32\%) of SNe Ia still belongs to this category ($X_{1} < 0.0$) even at $z > 0.5$ (see Figure 3 of \citealt{Kim2019}), assuming that all SNe Ia in cosmological sample will follow our relation between population age and SN Ia luminosity would require further justification. If high stretch ($X_{1} > 0.0$) SNe Ia in the cosmological sample somehow do not follow the same luminosity evolution, we would see a sizable difference in the Hubble residual between the low and high stretch subsamples. For the SNe Ia with diverse stretch values ($-3 < X_{1} < 2$), the analysis by \citet[][their Figure 3]{Scolnic2016}, however, shows only negligible difference in HR compared with the amount of the luminosity evolution suggested in this paper ($\Delta$HR $\sim$ 0.2 mag at $z > 0.5$), indicating that the luminosity evolution would have affected more or less equally to the cosmological sample regardless of the stretch factor. Only very high stretch ($X_{1} > 2$) subsample of \citet{Scolnic2016} shows some meaningful difference, but it comprises only $\sim$1.2\% of the total sample. \citet[][their Table~8]{Sullivan2011} also shows that the derived cosmological parameters are consistent between the low and high stretch SN Ia subsamples. Furthermore, a similar result we obtained for the luminosity evolution from the four different host properties in Table~\ref{table7}, which involves both low and high stretch SNe Ia, would also imply that our assumption in the cosmological analysis is not grossly incorrect. In Figure~\ref{figure17}, 1$\sigma$ significance interval of our evolution curve is added which is obtained from the regression fit of Figure~\ref{figure13}. Strikingly, the comparison of our evolution curves with SN data in Figures~\ref{figure16} and \ref{figure17} shows that the luminosity evolution can mimic a significant fraction of the HR (the difference in standardized SN brightness) used in the discovery and inference of the dark energy.\footnote{The detailed discussion on the inference of dark energy from other cosmological probes is beyond the scope of this paper. For the relevant discussion, we refer the reader to \citet{Tutusaus2017,Tutusaus2019} who showed that the low-$z$ probes and combination of them are consistent with a non-accelerating universe when the SN cosmology is highly affected by luminosity evolution (as suggested in this paper). When the cosmic microwave background (i.e., high-$z$ probe) data is added in their analysis, however, a nonaccelerated reconstruction is in tension with recent measurements of the Hubble constant.} This would indicate that the empirical standardization of SN Ia brightness practiced by the SNe Ia community is not quite capable of correcting for this effect of luminosity evolution. To put this result on a firmer refined basis, further observations (for more host sample, for the local property at the site of SN) are definitely required. Nevertheless, until then, this work presents some serious evidence for the luminosity evolution in SN cosmology. The future direction of SN cosmology, therefore, should investigate this systematic bias thoroughly before proceeding to the details of the dark energy.
\acknowledgments
We thank the anonymous referee for a number of helpful comments and suggestions. We also thank Nate Bastian, John Blakeslee, Aeree Chung, Pierre Demarque, Peter Garnavich, Myungkook J. Jee, Yong-Cheol Kim, Taysun Kimm, Pavel Kroupa, J. M. Diederik Kruijssen, Benjamin L'Huillier, Dongwook Lim, Mart\'in L\'opez-Corredoira, Myeong-Gu Park, Adam Riess, Arman Shafieloo, Elena Terlevich, Sukyoung K. Yi, and Suk-Jin Yoon for their comments and encouragements. We are grateful to the staff of LCO and MMT for their support during the observations. Support for this work was provided by the National Research Foundation of Korea (grants 2017R1A2B3002919, and 2017R1A5A1070354). Telescope times for this project were partially supported by K-GMT Science Program (PID: MMT\_2014\_00002,  GS-2015B-Q-41, GN-2016A-Q-51) of Korea Astronomy and Space Science Institute. Y. K. acknowledges support from the YONSEI University Graduate School Research Scholarship 2017. Y.-L.K. acknowledges support from the European Research Council (ERC) under the European Union's Horizon 2020 research and innovation programme (grant agreement No. 759194 - USNAC). This research has made use of the NASA/IPAC Extragalactic Database (NED) which is operated by the Jet Propulsion Laboratory, California Institute of Technology, under contract with the National Aeronautics and Space Administration. We acknowledge the usage of the HyperLeda (\href{http://leda.univ-lyon1.fr}{http://leda.univ-lyon1.fr}), SDSS (\href{http://www.sdss.org}{http://www.sdss.org}), $GALEX$ (\href{http://galex.stsci.edu}{http://galex.stsci.edu}), DESI Legacy Imaging Surveys (\href{http://legacysurvey.org}{http://legacysurvey.org}), and unWISE (\href{http://unwise.me}{http://unwise.me}) databases.

\clearpage

\clearpage

\clearpage

\clearpage

\clearpage

\clearpage

\clearpage

\clearpage

\clearpage

\clearpage

\clearpage

\clearpage

\clearpage

\clearpage

\clearpage

\clearpage

\begin{longrotatetable}
\begin{deluxetable*}{llccclclll}
\tablewidth{700pt}
\tabletypesize{\scriptsize}
\centering
\small
\tablecaption{\label{table1}Our Sample of Early-type Host Galaxies\tablenotemark{a}}
\tablehead{
SN         & Galaxy      &
\colhead{R.A.}          & \colhead{Decl.}  &
\colhead{${B}^{T}$}          & Morphology  &\colhead{$z$}    & PA   &Exposure   &\colhead{Telescope} \\
\colhead{}           & \colhead{}      &
\colhead{(J2000)}          & \colhead{(J2000)}  &
\colhead{(mag)}          & \colhead{}  &\colhead{}    &($^{\circ}$)    &($ N\times s$)   &\colhead{} \\
\vspace{-15pt}
}
\startdata
SN~1990af & 2MASX~J21345926-6244143 & 21:34:59.31 & -62:44:14.5 & 16.07 & SB0 & 0.050575 & 50 &$ 6 \times 3600$& LCO \\
SN~1992bo & ESO~352-G057 & 01:22:02.41 & -34:11:48.4 & 14.76 & SB(s)0\textasciicircum0\textasciicircum pec & 0.018983 & 12 &$3 \times 3600 $& LCO \\
SN~1993ac & CGCG~307-023 & 05:46:24.64 & +63:21:32.6 & 16.16 & E & 0.049000 & 135 &$3 \times 900$ & MMT \\
SN~1993ag & 2MASX~J10033546-3527410 & 10:03:35.46 & -35:27:41.0 & 16.86 & E3/S01 & 0.049094 & 80 &$11 \times 3600$& LCO \\
SN1994M & NGC~4493 & 12:31:08.37 & +00:36:49.3 & 14.78 & SA?0-pec & 0.023159 & 0 &$6 \times 3600 + 2 \times 600 + 2 \times 900$& LCO, MMT \\
SN~1997E & NGC~2258 & 06:47:45.80 & +74:28:54.0 & 13.01 & SA(r)0\textasciicircum0\textasciicircum & 0.013539 & 155 &$ 3 \times 900 $& MMT \\
SN~1997cn & NGC~5490 & 14:09:57.29 & +17:32:44.0 & 13.06 & E & 0.016195 & 5 &$2 \times 600 + 3 \times 900$& MMT \\
SN~1998bp & NGC~6495 & 17:54:50.76 & +18:19:36.9 & 13.85 & E & 0.010431 & 75 &$3 \times 600$& MMT \\
SN~1999ej & NGC~0495 & 01:22:56.00 & +33:28:18.0 & 13.68 & (R')SB(s)0/a & 0.013723 & 140 &$ 3 \times  900$& MMT \\
SN~2000dk & NGC~0382 & 01:07:23.87 & +32:24:13.9 & 14.48 & E:HII & 0.017442 & 135 &$3  \times 900$& MMT \\
SN~2001ie & UGC~05542 & 10:16:53.06 & +60:17:05.6 & 14.83 & E: & 0.030738 & 195 &$2 \times 600$& MMT \\
SN~2002G & CGCG~189-024 & 13:07:54.81 & +34:05:13.6 & 15.22 & E & 0.033737 & 75 &$2 \times 600$& MMT \\
SN~2002dj & NGC~5018 & 13:13:01.03 & -19:31:05.5 & 11.69 & E3: & 0.009393 & 100 &$3 \times 1800 + 4 \times 3600$& LCO \\
SN~2002do & MCG~+07-41-001 & 19:56:12.86 & +40:26:02.3 & 14.69 & E1: & 0.015881 & 140 &$2 \times 600$& MMT \\
SN~2002fb & NGC~0759 & 01:57:50.33 & +36:20:35.2 & 13.83 & E & 0.015567 & 70 &$3 \times 900$& MMT \\
SN~2003ch & UGC~03787 & 07:17:57.56 & +09:41:21.6 & 15.56 & E-S0 & 0.028620 & 0 &$ 3 \times 1800 + 1 \times 3600, 2 \times 900$ & LCO, MMT \\
SN~2003ic & MCG~-02-02-086 & 00:41:50.47 & -09:18:11.3 & 14.6 & cD;BrClG & 0.055672 & 130 &$6 \times 3600$& LCO \\
SN~2003iv & UGC~02320~NOTES01 & 02:50:08.84 & +12:50:37.0 & 15.79 & E\tablenotemark{\tiny b} & 0.034307 & 110 &$3 \times 900$& MMT \\
SN~2004gc & ARP327NED04 & 05:21:49.79 & +06:40:37.3 & 18.44 & SA0-: & 0.032089 & 160 &$2 \times 1200$& MMT \\
SN~2004gs & MCG~+03-22-020 & 08:38:23.85 & +17:37:52.8 & 15.21 & S0? & 0.027425 & 150 &$3\times 900$& MMT \\
SN~2005ag & 2MASX~J14564322+0919361   & 14:56:43.21 & +09:19:36.4 & 16.62 & SBbc\tablenotemark{\tiny c} & 0.079665 & 85 &$ 1 \times 1800$& MMT \\
SN~2005al & NGC~5304 & 13:50:01.48 & -30:34:42.5 & 13.62 & E+pec: & 0.012402 & 140 &$ 5 \times 3600$& LCO\\
SN~2005bl& NGC~4070 & 12:04:11.30 & +20:24:35.4 & 14.06 & E & 0.024060 & 170 &$2 \times 600$& MMT \\
SN~2005el & NGC~1819 & 05:11:46.14 & +05:12:02.2 & 13.69 & SB0;Sbrst & 0.014910 & 130 &$1 \times 1800, 3 \times 900$& LCO, MMT \\
SN~2005mc & UGC~04414 & 08:27:05.96 & +21:38:43.0 & 14.41 & S0a & 0.025221 & 120 &$3 \times 900$& MMT \\
SN~2006N & CGCG~308-009 & 06:08:30.39 & +64:43:25.3 & 14.6 & E\tablenotemark{\tiny b} & 0.014277 & 80 &$1 \times  900$& MMT \\
SN~2006bd & UGC~06609 & 11:38:29.46 & +20:31:39.8 & 14.72 & E & 0.025724 & 100 &$2 \times 600 + 2 \times 900$& MMT \\
SN~2006bq & NGC~6685 & 18:39:58.64 & +39:58:54.4 & 14.59 & S0-: & 0.021905 & 20 &$2 \times 600$& MMT \\
SN~2006cs & CGCG~190-050 & 13:45:34.99 & +35:36:40.1 & 14.77 & S0? & 0.023676 & 34 &$2 \times  600$& MMT \\
SN~2006ef & NGC~0809 & 02:04:18.97 & -08:44:07.1 & 14.59 & (R)S0+: & 0.017902 & 175 &$3 \times  3600$& LCO  \\
SN~2006ej & IC~1563 & 00:39:00.24 & -09:00:52.5 & 14.59 & S0pecsp & 0.020452 & 90 &$1 \times  1800$& LCO  \\
SN~2006kf & UGC~02829 & 03:41:50.86 & +08:09:35.4 & 14.94 & S0 & 0.021301 & 160 & $4 \times 1800, 1 \times 900$& LCO, MMT \\
SN~2006ot & ESO~544-G031 & 02:15:04.60 & -20:46:03.7 & 15.4 & Sa & 0.053100 & 4 &$1 \times 3600$& LCO \\
SN~2007R & MCG~+08-14-043 & 07:46:37.71 & +44:47:25.9 & 14.38 & S0/a & 0.030880 & 165 &$3 \times 900$& MMT \\
SN~2007ap & MCG~+03-41-003 & 15:56:23.59 & +16:31:23.9 & 14.65 & SA0+ & 0.015818 & 40 &$2 \times600$& MMT \\
SN2007au & UGC~03725 & 07:11:41.82 & +49:52:00.0 & 14.15 & S0-: & 0.020584 & 125 &$3\times 900$& MMT \\
SN~2007ci & NGC~3873 & 11:45:46.10 & +19:46:26.2 & 13.85 & E & 0.018126 & 0 &$2 \times 600$& MMT \\
SN~2007cp & IC~0807 & 12:42:12.49 & -17:24:12.8 & 14.6 & E-S0\tablenotemark{\tiny b} & 0.036595 & 95 &$1 \times 3600$& LCO  \\
SN~2007hj & NGC~7461 & 23:01:48.33 & +15:34:56.9 & 14.48 & SB0 & 0.014113 & 155 &$2 \times  900$& MMT \\
SN~2007jh  & CGCG~391-014 & 03:36:01.59 & +01:06:17.1 & 15.51 & E0 & 0.040796 & 55 &$3 \times  900$& MMT \\
SN~2007nq & UGC~00595 & 00:57:34.92 & -01:23:27.9 & 14.38 & E;Radio galaxy;LEG & 0.045031 & 140 &$3 \times 3600$& LCO  \\
SN~2008C & UGC~03611 & 06:57:11.73 & +20:26:14.5 & 15.05 & S0/a & 0.016621 & 60 &$3 \times 900$& MMT \\
SN~2008L & NGC~1259 & 03:17:17.28 & +41:23:07.8 & 15.21 & S0 & 0.019400 & 100/10 &$1 \times 900/1 \times 900$& MMT \\
SN~2008R & NGC~1200 & 03:03:54.48 & -11:59:30.5 & 13.8 & SA(s)0- & 0.013503 & 130 &$3 \times  3600$& LCO  \\
SN~2008af & UGC~09640 & 14:59:27.64 & +16:38:42.4 & 14.32 & E & 0.033507 & 100 &$2 \times 600$& MMT \\
SN~2008gl & UGC~00881 & 01:20:53.47 & +04:48:04.8 & 15.26 & E & 0.034017 & 40 &$3 \times 900$& MMT \\
SN~2008hv & NGC~2765 & 09:07:36.64 & +03:23:34.5 & 13.18 & S0 & 0.012549 & 107 &$2 \times 1800 + 5 \times 3600, 4 \times 900$& LCO, MMT \\
SN~2008ia & ESO~125-G006 & 08:50:35.85 & -61:16:40.5 & 11.94$I$ & S0 & 0.021942 & 35 &$6 \times 3600$& LCO \\
SN~2009F  & NGC~1725 & 04:59:22.89 & -11:07:56.3 & 13.82 & S0 & 0.012956 & 15 &$3 \times 3600$& LCO \\
SN~2010H  & IC~0494 & 08:06:24.12 & +01:02:09.8 & 14.26 & SA0\textasciicircum0\textasciicircum: & 0.015367 & 45 &$3 \times 900$& MMT \\
SN~2010Y & NGC~3392 & 10:51:03.00 & +65:46:53.6 & 14.59 & E?BLAGN & 0.010860 & 115 &$ 2 \times 900$& MMT \\
\enddata
\tablenotetext{a}{Information from NED and HyperLeda (the total $B$ magnitude, $B^{T}$).}
\tablenotetext{b}{Morphological classification is adopted from HyperLeda.}
\tablenotetext{c}{Initially classified as an early-type.}
\end{deluxetable*}
\end{longrotatetable}

\clearpage

\clearpage
\startlongtable
\begin{deluxetable*}{llcccccccc}
\tablewidth{700pt}
\tabletypesize{\scriptsize}
\centering
\small
\tablecaption{SALT2 Light-curve Parameters of SNe Ia in Our Sample\label{table2}}
\tablehead{\vspace{-7pt}\\
SN & Galaxy
  & \colhead{$m_{B}$}   & \colhead{Error} 
  & \colhead{$X_{1}$}   & \colhead{Error} 
  & \colhead{$C$}   & \colhead{Error} 
  & \colhead{HR}   & \colhead{Error}  \\
   \colhead{}   & \colhead{}
 &\colhead{(mag)}   & \colhead{} 
 & \colhead{}   & \colhead{} 
 & \colhead{}   & \colhead{} 
 & \colhead{(mag)}   & \colhead{}  
}
\startdata
SN~1990af & 2MASX~J21345926-6244143 & 17.777 & 0.032 & -2.132 & 0.141 & -0.043 & 0.044 & -0.068 & 0.167  \\
SN~1992bo & ESO~352-G057 & 15.782 & 0.033 & -1.98 & 0.064 & -0.045 & 0.028 & 0.234 & 0.109  \\
SN~1993ac & CGCG~307-023 & 17.777 & 0.108 & -0.896 & 0.521 & 0.016 & 0.042 & -0.059 & 0.204  \\
SN~1993ag & 2MASX~J10033546-3527410 & 17.834 & 0.034 & -0.832 & 0.136 & 0.057 & 0.045 & -0.176 & 0.171  \\
SN~1994M & NGC~4493 & 16.294 & 0.035 & -1.405 & 0.090 & 0.055 & 0.028 & -0.186 & 0.110   \\
SN~1997E & NGC~2258 & 15.114 & 0.029 & -1.702 & 0.087 & 0.016 & 0.022 & 0.060 & 0.087  \\
SN~1997cn & NGC~5490 & \nodata & \nodata & \nodata &\nodata & \nodata & \nodata & \nodata  & \nodata \\
SN~1998bp & NGC~6495 & 15.308 & 0.034 & -2.492 & 0.092 & 0.194 & 0.024 & 0.06 & 0.096   \\
SN~1999ej & NGC~0495 & 15.339 & 0.052 & -1.485 & 0.223 & -0.032 & 0.031 & 0.578 & 0.13  \\
SN~2000dk & NGC~0382 & 15.364 & 0.027 & -2.047 & 0.077 & -0.023 & 0.022 & -0.06 & 0.086\\
SN~2001ie & UGC~05542 & 16.666 & 0.042 & -0.588 & 0.119 & -0.024 & 0.027 & 0.046 & 0.110 \\
SN~2002G & CGCG~189-024 & 18.450 & 0.035 & 0.197 & 0.050 & -0.046 & 0.023 & -0.265 & 0.172  \\
SN~2002dj & NGC~5018 & 13.955 & 0.033 & 0.006 & 0.129 & 0.063 & 0.023 & -0.478 & 0.093 \\
SN~2002do & MCG~+07-41-001 & 15.473 & 0.032 & -3.04 & 0.178 & -0.044 & 0.027 & 0.146 & 0.108 \\
SN~2002fb & NGC~0759 & \nodata & \nodata & \nodata & \nodata & \nodata & \nodata & \nodata & \nodata  \\
SN~2003ch & UGC~03787 & 16.675 & 0.028 & -1.347 & 0.155 & -0.039 & 0.023 & 0.435 & 0.092  \\
SN~2003ic~ & MCG~-02-02-086 & 17.607 & 0.03 & -1.999 & 0.203 & -0.066 & 0.029 & -0.306 & 0.115 \\
SN~2003iv & UGC~ 02320 NOTES01 & 16.973 & 0.028 & -2.157 & 0.188 & -0.095 & 0.027 & 0.221 & 0.107  \\
SN~2004gc & ARP~327NED04 & 16.721 & 0.076 & -0.682 & 0.04 & 0.078 & 0.022 & -0.254 & 0.111   \\
SN~2004gs & MCG~+03-22-020 & 17.140 & 0.082 & -1.844 & 0.032 & 0.121 & 0.019 & 0.076 & 0.108  \\
SN~2005ag & SDSS J145643.21+091936.4  & 18.450 & 0.035 & 0.197 & 0.050 & -0.046 & 0.023 & -0.095 & 0.092  \\
SN~2005al & NGC~5304 & 14.865 & 0.165 & -1.220 & 0.029 & -0.096 & 0.019 & 0.296 & 0.179  \\
SN~2005bl & NGC~4070 & \nodata & \nodata & \nodata & \nodata & \nodata & \nodata & \nodata & \nodata \\
SN~2005el & NGC~1819 & 14.843 & 0.028 & -1.342 & 0.032 & -0.131 & 0.021 & 0.151 & 0.083 \\
SN~2005mc & UGC~04414 & 17.031 & 0.089 & -1.916 & 0.044 & 0.186 & 0.021 & -0.159 & 0.118  \\
SN~2006N & CGCG~308-009 & 15.055 & 0.029 & -1.955 & 0.059 & -0.097 & 0.021 & 0.221 & 0.083   \\
SN~2006bd & UGC~06609 & \nodata & \nodata & \nodata & \nodata & \nodata & \nodata  & \nodata & \nodata \\
SN~2006bq & NGC~6685 & 16.193 & 0.027 & -1.568 & 0.059 & 0.028 & 0.022 & 0.058 & 0.086 \\
SN~2006cs & CGCG~190-050 & \nodata & \nodata & \nodata & \nodata & \nodata & \nodata & \nodata & \nodata \\
SN~2006ef & NGC~0809 & 15.443 & 0.083 & -1.092 & 0.204 & -0.094 & 0.042 & 0.347 & 0.178 \\
SN~2006ej & IC~1563 & 15.680 & 0.026 & -1.488 & 0.041 & -0.041 & 0.020 & 0.061 & 0.078  \\
SN~2006kf & UGC~02829 & 15.805 & 0.107 & -1.993 & 0.054 & -0.079 & 0.020 & 0.074 & 0.130 \\
SN~2006ot & ESO~544-G031 & 17.982 & 0.047 & 0.641 & 0.086 & 0.093 & 0.020 & -0.037 & 0.088 \\
SN~2007R & MCG~+08-14-043 & 16.697 & 0.045 & -1.346 & 0.105 & -0.125 & 0.031 & 0.336 & 0.124 \\
SN~2007ap & MCG~+03-41-003 & 15.348 & 0.041 & -1.688 & 0.069 & -0.109 & 0.029 & 0.324 & 0.115   \\
SN~2007au & UGC~03725 & 16.550 & 0.033 & -3.090 & 0.116 & 0.144 & 0.028 & -0.104 & 0.110  \\
SN~2007ci & NGC~3873 & 15.895 & 0.032 & -2.827 & 0.099 & 0.011 & 0.027 & -0.105 & 0.106  \\
SN~2007cp & IC~0807 & 17.133 & 0.040 & -4.011 & 0.88 & -0.052 & 0.040 & -0.319 & 0.202   \\
SN~2007hj & NGC~7461 & 15.520 & 0.035 & -2.137 & 0.063 & 0.100 & 0.027 & 0.157 & 0.106 \\
SN~2007jh & CGCG~391-014 & \nodata & \nodata & \nodata & \nodata & \nodata & \nodata & \nodata & \nodata \\
SN~2007nq & UGC~00595 & 17.414 & 0.054 & -1.917 & 0.051 & -0.028 & 0.02 & -0.153 & 0.092  \\
SN~2008C & UGC~03611 & 15.644 & 0.129 & -0.719 & 0.046 & 0.126 & 0.019 & -0.221 & 0.147  \\
SN~2008L & NGC~1259 & 15.097 & 0.037 & -1.592 & 0.100 & -0.151 & 0.032 & -0.097 & 0.125 \\
SN~2008R & NGC~1200 & 15.252 & 0.170 & -2.170 & 0.048 & 0.047 & 0.020 & 0.080 & 0.185 \\
SN~2008af & UGC~09640 & 16.857 & 0.057 & -1.425 & 0.130 & -0.031 & 0.032 & -0.061 & 0.133   \\
SN~2008gl & UGC~00881 & 16.792 & 0.03 & -1.408 & 0.098 & -0.04 & 0.028 & -0.017 & 0.109  \\
SN~2008hv & NGC~2765 & 14.732 & 0.162 & -1.220 & 0.030 & -0.068 & 0.019 & 0.011 & 0.177  \\
SN~2008ia & ESO125-G006 & 15.841 & 0.099 & -1.312 & 0.044 & -0.039 & 0.019 & -0.118 & 0.121 \\
SN~2009F & NGC~1725 & \nodata & \nodata & \nodata & \nodata & \nodata & \nodata & \nodata & \nodata \\
SN~2010H & IC~0494 & \nodata & \nodata & \nodata & \nodata & \nodata & \nodata & \nodata & \nodata \\
SN~2010Y & NGC~3392 & 14.987 & 0.040 & -2.503 & 0.084 & -0.07 & 0.031 & 0.487 & 0.122 \\
\enddata
\end{deluxetable*}
\clearpage

\begin{deluxetable*}{p{5cm}lp{5cm}l}
\center
\caption{Instrumental Setup for Observations at MMT 6.5 m\label{table3}}
\tablehead{\multicolumn{2}{c}{}\vspace{-5pt}\\
\multicolumn{2}{c}{Blue Channel Spectrograph}\vspace{3pt}}
\startdata
Grating and Blaze & 300 l mm$^{-1}$ , 5500 {\AA} \\
Spectral range & 3200-8000 {\AA} \\
Resolution & $\sim$8 {\AA} /FWHM \\
Dispersion & 1.96{\AA}/pixel \\
Slit width  & 1{\arcsec}.5 \\
Slit length & 180{\arcsec}\\
\enddata
\tablecomments{Dates of observations: 2014 May 1-2; 2015 January 15 - 17; 2016 December 22-23, 25-27}
\end{deluxetable*}
\clearpage

\startlongtable
\begin{deluxetable*}{llcccccccccc}
\tablewidth{700pt}
\tabletypesize{\scriptsize}
\centering
\small
\tablecaption{Line Measurements of Our Host Galaxy Sample\label{table4}}
\tablehead{\vspace{-7pt}\\
SN & Galaxy 
  & \colhead{$\sigma$$_{\textit{v}}$}   & \colhead{Error} 
  & \colhead{H$\beta$}   & \colhead{Error} 
  & \colhead{Mg\,$b$}   & \colhead{Error} 
  & \colhead{Fe5270}   & \colhead{Error} 
  & \colhead{Fe5335}   & \colhead{Error}\\
   \colhead{}   & \colhead{}
 &\colhead{(km s$^{-1}$)}   & \colhead{} 
 & \colhead{(\AA)}   & \colhead{} 
 & \colhead{(\AA)}   & \colhead{} 
 & \colhead{(\AA)}   & \colhead{}  
 & \colhead{(\AA)}   & \colhead{}
}
\startdata
SN~1990af & 2MASX~J21345926-6244143 & 191.7 & 3.3 & 1.683 & 0.060 & 4.255 & 0.065 & 3.263 & 0.148 & 3.186 & 0.113 \\
SN~1992bo & ESO~352-~G~057 & 180.3 & 2.7 & 1.620 & 0.038 & 4.248 & 0.040 & 2.945 & 0.043 & 3.052 & 0.051 \\
SN~1993ac & CGCG~307-023 & 284.1 & 5.8 & 1.615 & 0.033 & 4.944 & 0.042 & 2.861 & 0.045 & 2.827 & 0.068 \\
SN~1993ag & 2MASX~J10033546-3527410 & 201.7 & 2.8 & 1.604 & 0.034 & 3.911 & 0.036 & 2.754 & 0.121 & 2.709 & 0.092 \\
SN~1994M & NGC~4493 & 220.6 & 2.9 & 1.744 & 0.044 & 5.083 & 0.048 & 2.930 & 0.052 & 2.850 & 0.064 \\
SN~1997E & NGC~2258 & 302.0 & 9.6 & 1.566 & 0.019 & 5.455 & 0.022 & 3.208 & 0.024 & 2.870 & 0.033 \\
SN~1997cn & NGC~5490 & 356.2 & 5.1 & 1.480 & 0.017 & 5.464 & 0.021 & 3.181 & 0.023 & 2.983 & 0.033 \\
SN~1998bp & NGC~6495 & 198.3 & 12.7 & 1.561 & 0.037 & 5.474 & 0.039 & 3.018 & 0.045 & 2.679 & 0.056 \\
SN~1999ej & NGC~0495 & 157.4 & 15.1 & 1.966 & 0.032 & 4.459 & 0.035 & 3.294 & 0.038 & 3.086 & 0.045 \\
SN~2000dk & NGC~0382 & 196.1 & 11.9 & 1.563 & 0.033 & 4.535 & 0.038 & 3.358 & 0.042 & 3.016 & 0.050 \\
SN~2001ie & UGC~05542 & 205.5 & 5.5 & 1.683 & 0.060 & 3.577 & 0.068 & 2.951 & 0.072 & 2.784 & 0.086 \\
SN~2002G & CGCG~189-024 & 142.0 & 9.4 & 1.827 & 0.047 & 4.978 & 0.053 & 2.978 & 0.059 & 2.794 & 0.074 \\
SN~2002dj & NGC5018 & 215.4 & 2.4 & 2.618 & 0.009 & 3.229 & 0.009 & 2.781 & 0.010 & 2.570 & 0.013 \\
SN~2002do & MCG+07-41-001 & 288.6 & 9.0 & 1.627 & 0.015 & 5.282 & 0.018 & 3.106 & 0.020 & 3.153 & 0.027 \\
SN~2002fb & NGC~0759 & 272.0 & 12.3 & 2.814 & 0.026 & 3.637 & 0.031 & 2.648 & 0.033 & 2.520 & 0.043 \\
SN~2003ch & UGC03787 & 199.6 & 3.3 & 2.610 & 0.052 & 3.495 & 0.054 & 2.605 & 0.059 & 2.429 & 0.072 \\
SN~2003ic & MCG-02-02-086 & 350.2 & 5.7 & 1.380 & 0.064 & 5.175 & 0.077 & 3.054 & 0.148 & 2.582 & 0.124 \\
SN~2003iv & UGC~02320~NOTES01 & 206.9 & 12.1 & 1.811 & 0.030 & 4.417 & 0.034 & 3.023 & 0.039 & 2.761 & 0.047 \\
SN~2004gc & ARP327NED04 & 20.7 & 64.6 & 4.710 & 0.047 & 2.776 & 0.051 & 1.746 & 0.059 & 1.796 & 0.066 \\
SN~2004gs & MCG+03-22-020 & 115.6 & 21.3 & 2.018 & 0.044 & 4.039 & 0.047 & 2.959 & 0.051 & 2.666 & 0.058 \\
SN~2005ag & 2MASX~J14564322+0919361  & 172.7 & 9.4 & 4.982 & 0.104 & 2.256 & 0.144 & 2.017 & 0.125 & 1.680 & 0.144 \\
SN~2005al & NGC5304 & 222.5 & 4.4 & 1.836 & 0.034 & 4.476 & 0.035 & 2.785 & 0.039 & 2.719 & 0.049 \\
SN~2005bl & NGC4070 & 175.1 & 10.8 & 2.000 & 0.045 & 3.955 & 0.050 & 2.914 & 0.054 & 2.872 & 0.066 \\
SN~2005el & NGC1819 & 195.1 & 7.1 & 4.548 & 0.081 & 2.913 & 0.082 & 2.242 & 0.093 & 2.032 & 0.121 \\
SN~2005mc & UGC04414 & 216.9 & 11.6 & 1.454 & 0.044 & 4.834 & 0.047 & 3.129 & 0.053 & 2.768 & 0.066 \\
SN~2006N & CGCG308-009 & 167.1 & 4.5 & 1.968 & 0.042 & 4.373 & 0.045 & 2.792 & 0.051 & 2.579 & 0.063 \\
SN~2006bd & UGC06609 & 211.2 & 7.9 & 1.636 & 0.037 & 5.168 & 0.041 & 2.928 & 0.046 & 2.700 & 0.057 \\
SN~2006bq & NGC6685 & 175.0 & 13.1 & 1.782 & 0.043 & 4.951 & 0.047 & 2.845 & 0.053 & 2.688 & 0.066 \\
SN~2006cs & CGCG190-050 & 227.4 & 6.3 & 1.841 & 0.040 & 4.512 & 0.045 & 3.007 & 0.050 & 2.644 & 0.061 \\
SN~2006ef & NGC0809 & 180.6 & 3.0 & 2.014 & 0.035 & 4.234 & 0.036 & 3.158 & 0.040 & 3.168 & 0.049 \\
SN~2006ej & IC1563 & 233.1 & 4.9 & 1.828 & 0.140 & 4.903 & 0.153 & 2.562 & 0.171 & 2.835 & 0.221 \\
SN~2006kf & UGC02829 & 284.1 & 4.4 & 1.454 & 0.029 & 4.648 & 0.032 & 3.017 & 0.035 & 3.014 & 0.046 \\
SN~2006ot & ESO544-G031 & 286.7 & 3.9 & 1.764 & 0.089 & 4.681 & 0.102 & 3.149 & 0.114 & 2.768 & 0.148 \\
SN~2007R & MCG+08-14-043 & 211.1 & 9.4 & 1.663 & 0.032 & 4.360 & 0.037 & 2.980 & 0.042 & 2.899 & 0.049 \\
SN~2007ap & MCG+03-41-003 & 105.9 & 17.5 & 2.081 & 0.044 & 4.398 & 0.045 & 3.144 & 0.050 & 2.912 & 0.057 \\
SN~2007au & UGC03725 & 290.9 & 9.5 & 1.453 & 0.023 & 5.766 & 0.029 & 3.017 & 0.032 & 2.761 & 0.041 \\
SN~2007ci & NGC3873 & 244.6 & 5.0 & 1.826 & 0.027 & 4.678 & 0.031 & 2.830 & 0.035 & 2.721 & 0.044 \\
SN~2007cp & IC0807 & 189.1 & 4.8 & 2.232 & 0.077 & 4.054 & 0.080 & 2.720 & 0.090 & 2.562 & 0.110 \\
SN~2007hj & NGC7461 & 154.8 & 4.5 & 1.632 & 0.029 & 4.969 & 0.031 & 2.817 & 0.034 & 2.610 & 0.041 \\
SN~2007jh & CGCG391-014 & 198.4 & 4.1 & 1.910 & 0.035 & 4.391 & 0.041 & 2.990 & 0.046 & 2.602 & 0.060 \\
SN~2007nq & UGC00595 & 244.4 & 4.3 & 1.668 & 0.073 & 4.940 & 0.079 & 2.851 & 0.088 & 2.955 & 0.160 \\
SN~2008C & UGC03611 & 74.7 & 23.9 & 4.403 & 0.033 & 2.791 & 0.034 & 2.315 & 0.038 & 1.817 & 0.044 \\
SN~2008L & NGC1259 & 147.5 & 5.0 & 2.197 & 0.049 & 4.406 & 0.053 & 2.748 & 0.060 & 2.400 & 0.073 \\
SN~2008R & NGC1200 & 249.8 & 3.7 & 1.652 & 0.031 & 5.103 & 0.034 & 3.239 & 0.037 & 3.072 & 0.048 \\
SN~2008af & UGC09640 & 218.6 & 5.8 & 1.868 & 0.045 & 4.725 & 0.051 & 3.250 & 0.056 & 2.950 & 0.069 \\
SN~2008gl & UGC00881 & 177.2 & 4.3 & 2.087 & 0.063 & 4.689 & 0.071 & 2.891 & 0.079 & 2.905 & 0.095 \\
SN~2008hv & NGC2765 & 190.7 & 2.5 & 1.897 & 0.024 & 3.989 & 0.025 & 2.983 & 0.028 & 2.792 & 0.034 \\
SN~2008ia & ESO125-G006 & 213.0 & 2.4 & 1.920 & 0.013 & 4.036 & 0.014 & 2.913 & 0.016 & 2.731 & 0.020 \\
SN~2009F & NGC1725 & 273.8 & 4.6 & 1.789 & 0.028 & 5.328 & 0.030 & 3.096 & 0.033 & 3.161 & 0.043 \\
SN~2010H & IC0494 & 91.3 & 5.7 & 2.771 & 0.026 & 3.368 & 0.029 & 2.986 & 0.031 & 2.778 & 0.036 \\
SN~2010Y & NGC3392 & 121.8 & 5.2 & 2.089 & 0.047 & 4.556 & 0.049 & 2.745 & 0.054 & 2.437 & 0.064 \\
\enddata
\end{deluxetable*}
\clearpage

\startlongtable
\begin{deluxetable*}{ll@{\extracolsep{\fill}}CCCCCC}
\tablewidth{700pt}
\tabletypesize{\scriptsize}
\centering
\small
\tablecaption{Determined Age and Metallicity of Our Host Galaxy Sample\label{table5}}
\tablehead{\vspace{-7pt}\\
{SN} & {Galaxy}  
  & \multicolumn{2}{c}{YEPS} & \multicolumn{2}{c}{TMJ11} 
  &  \multicolumn{2}{c}{S07}\\
  \cline{3-4} \cline{5-6} \cline{7-8}
 \colhead{}   & \colhead{} 
  & \colhead{Age}   & \colhead{[M/H]} 
  & \colhead{Age}   & \colhead{[M/H]} 
   & \colhead{Age}   & \colhead{[M/H]}\\[-0.5em]
    \colhead{}   & \colhead{} 
 & \colhead{(Gyr)}   & \colhead{(dex)} 
  & \colhead{(Gyr)}   & \colhead{(dex)} 
   & \colhead{(Gyr)}   & \colhead{(dex)} 
}
\startdata
SN~1990af & 2MASX~J21345926-6244143 & 5.38^{+1.40}_{-0.78} & 0.39^{+0.15}_{-0.14} & 9.20^{+2.12}_{-2.80} & 0.37^{+0.11}_{-0.13} & 9.82^{+1.71}_{-1.51} & 0.15^{+0.09}_{-0.09} \\
SN~1992bo & ESO~352-G057 & 7.17^{+0.97}_{-0.84} & 0.21^{+0.06}_{-0.06} & 11.87^{+1.23}_{-1.11} & 0.17^{+0.05}_{-0.05} & 11.80^{+0.88}_{-0.99} & 0.07^{+0.04}_{-0.05} \\
SN~1993ac & CGCG~307-023 & 6.65^{+0.83}_{-0.73} & 0.35^{+0.07}_{-0.06} & 12.47^{+1.24}_{-1.17} & 0.22^{+0.06}_{-0.06} & 11.76^{+0.87}_{-0.83} & 0.20^{+0.05}_{-0.06} \\
SN~1993ag & 2MASX~J10033546-3527410 & 9.28^{+1.20}_{-1.19} & -0.03^{+0.11}_{-0.11} & 12.93^{+1.57}_{-2.29} & 0.33^{+0.09}_{-0.10} & 12.92^{+1.37}_{-1.13} & -0.07^{+0.09}_{-0.09} \\
SN~1994M & NGC~4493 & 4.77^{+0.65}_{-0.43} & 0.51^{+0.08}_{-0.07} & 8.91^{+1.49}_{-1.68} & 0.38^{+0.06}_{-0.07} & 9.11^{+1.07}_{-0.88} & 0.30^{+0.05}_{-0.05} \\
SN~1997E & NGC~5490 & 5.69^{+0.32}_{-0.28} & 0.62^{+0.03}_{-0.03} & 12.28^{+0.99}_{-0.98} & 0.38^{+0.03}_{-0.03} & 11.71^{+0.48}_{-0.46} & 0.33^{+0.02}_{-0.02} \\
SN~1997cn & NGC~2258 & 6.80^{+0.40}_{-0.36} & 0.58^{+0.04}_{-0.04} & 14.95^{+0.69}_{-1.05} & 0.34^{+0.03}_{-0.03} & 13.50^{+0.63}_{-0.60} & 0.30^{+0.02}_{-0.03} \\
SN~1998bp & NGC~6495 & 6.37^{+0.72}_{-0.69} & 0.51^{+0.06}_{-0.06} & 13.78^{+1.39}_{-1.30} & 0.26^{+0.06}_{-0.06} & 12.45^{+1.20}_{-0.89} & 0.30^{+0.04}_{-0.06} \\
SN~1999ej & NGC~0495 & 3.46^{+0.21}_{-0.24} & 0.66^{+0.06}_{-0.06} & 3.52^{+0.39}_{-0.38} & 0.55^{+0.05}_{-0.04} & 5.62^{+0.48}_{-0.40} & 0.31^{+0.04}_{-0.04} \\
SN~2000dk & NGC~0382 & 6.68^{+0.86}_{-0.70} & 0.36^{+0.06}_{-0.06} & 12.40^{+1.20}_{-1.09} & 0.26^{+0.05}_{-0.05} & 12.10^{+0.95}_{-0.87} & 0.15^{+0.04}_{-0.04} \\
SN~2001ie & UGC~05542 & 7.78^{+1.44}_{-1.44} & 0.03^{+0.09}_{-0.10} & 11.21^{+1.83}_{-1.61} & 0.01^{+0.08}_{-0.10} & 11.11^{+1.49}_{-1.39} & -0.07^{+0.08}_{-0.08} \\
SN~2002G & CGCG~189-024 & 4.29^{+0.51}_{-0.44} & 0.53^{+0.09}_{-0.09} & 6.95^{+1.78}_{-1.54} & 0.40^{+0.07}_{-0.07} & 7.90^{+0.93}_{-0.89} & 0.30^{+0.05}_{-0.05} \\
SN~2002dj & NGC~5018 & 2.11^{+0.06}_{-0.04} & 0.45^{+0.02}_{-0.02} & 1.79^{+0.01}_{-0.01} & 0.41^{+0.01}_{-0.01} & 2.32^{+0.04}_{-0.03} & 0.23^{+0.02}_{-0.02} \\
SN~2002do & MCG~+07-41-001 & 5.03^{+0.20}_{-0.16} & 0.63^{+0.03}_{-0.02} & 9.85^{+0.65}_{-0.54} & 0.43^{+0.02}_{-0.02} & 10.42^{+0.39}_{-0.38} & 0.33^{+0.02}_{-0.02} \\
SN~2002fb & NGC~0759 & 1.78^{+0.04}_{-0.10} & 0.58^{+0.08}_{-0.03} & 1.62^{+0.03}_{-0.03} & 0.62^{+0.05}_{-0.04} & 1.92^{+0.09}_{-0.13} & 0.36^{+0.04}_{-0.04} \\
SN~2003ch & UGC~03787 & 2.21^{+0.25}_{-0.19} & 0.39^{+0.09}_{-0.09} & 1.83^{+0.07}_{-0.06} & 0.43^{+0.07}_{-0.08} & 2.44^{+0.18}_{-0.19} & 0.22^{+0.06}_{-0.06} \\
SN~2003ic & MCG-02-02-086 & 10.48^{+2.31}_{-2.16} & 0.27^{+0.15}_{-0.17} & 19.60^{+2.51}_{-2.75} & 0.07^{+0.14}_{-0.16} & 18.01^{+3.60}_{-2.95} & 0.14^{+0.13}_{-0.12} \\
SN~2003iv & PGC~010738 & 4.85^{+0.46}_{-0.34} & 0.35^{+0.06}_{-0.05} & 7.78^{+0.98}_{-0.91} & 0.31^{+0.05}_{-0.05} & 8.37^{+0.62}_{-0.60} & 0.17^{+0.03}_{-0.03} \\
SN~2004gc & ARP~327NED04 & 0.10^{+ \scalebox{0.6}{\textless}0.01}_{-\scalebox{0.6}{\textless}0.01} & 0.32^{+0.06}_{-0.03} & 0.64^{+0.01}_{-0.09} & 0.97^{+0.02}_{-0.09} & 1.10^{+ \scalebox{0.6}{\textless}0.01}_{-\scalebox{0.6}{\textless}0.01} & -0.08^{+0.04}_{-0.03} \\
SN~2004gs & MCG+03-22-020 & 3.90^{+0.40}_{-0.28} & 0.32^{+0.08}_{-0.08} & 4.65^{+0.70}_{-0.80} & 0.35^{+0.06}_{-0.07} & 5.67^{+0.85}_{-0.52} & 0.16^{+0.05}_{-0.04} \\
SN~2005ag & SDSS~J145643.21+091936.4 & 0.10^{+ \scalebox{0.6}{\textless}0.01}_{-\scalebox{0.6}{\textless}0.01} & 0.35^{+0.05}_{-0.03} & 0.54^{+0.05}_{-0.02} & 0.52^{+0.37}_{-0.24} & 1.10^{+\scalebox{0.6}{\textless}0.01}_{-\scalebox{0.6}{\textless}0.01} & -0.25^{+0.08}_{-0.07} \\
SN~2005al & NGC~5304 & 5.00^{+0.55}_{-0.40} & 0.29^{+0.06}_{-0.05} & 7.86^{+1.00}_{-0.98} & 0.27^{+0.05}_{-0.05} & 8.31^{+0.59}_{-0.61} & 0.18^{+0.04}_{-0.05} \\
SN~2005bl & NGC~4070 & 3.92^{+0.41}_{-0.28} & 0.34^{+0.09}_{-0.08} & 4.69^{+0.74}_{-0.85} & 0.35^{+0.06}_{-0.07} & 5.81^{+0.81}_{-0.59} & 0.16^{+0.05}_{-0.04} \\
SN~2005el & NGC~1819 & 0.10^{+ \scalebox{0.6}{\textless}0.01}_{-\scalebox{0.6}{\textless}0.01} & 0.49^{+0.09}_{-0.03} & 0.63^{+0.03}_{-0.02} & 0.99^{+\scalebox{0.6}{\textless}0.01}_{-\scalebox{0.6}{\textless}0.01} & 1.09^{+0.01}_{-0.01} & -0.03^{+0.46}_{-0.32} \\
SN~2005mc & UGC~04414 & 9.13^{+1.26}_{-1.21} & 0.27^{+0.08}_{-0.07} & 16.75^{+1.65}_{-1.69} & 0.15^{+0.07}_{-0.07} & 15.44^{+1.76}_{-1.51} & 0.13^{+0.06}_{-0.06} \\
SN~2006N & CGCG~308-009 & 4.27^{+0.47}_{-0.40} & 0.31^{+0.08}_{-0.07} & 5.52^{+0.89}_{-0.80} & 0.31^{+0.06}_{-0.07} & 6.67^{+0.63}_{-0.78} & 0.19^{+0.04}_{-0.06} \\
SN~2006bd & UGC~06609 & 6.10^{+0.71}_{-0.68} & 0.43^{+0.06}_{-0.06} & 11.98^{+1.25}_{-1.14} & 0.25^{+0.06}_{-0.06} & 11.32^{+0.86}_{-0.89} & 0.25^{+0.05}_{-0.06} \\
SN~2006bq & NGC~6685 & 4.93^{+0.71}_{-0.47} & 0.41^{+0.08}_{-0.07} & 8.93^{+1.20}_{-1.54} & 0.31^{+0.06}_{-0.07} & 8.91^{+0.95}_{-0.81} & 0.25^{+0.05}_{-0.07} \\
SN~2006cs & CGCG~190-050 & 4.73^{+0.57}_{-0.43} & 0.35^{+0.07}_{-0.07} & 7.39^{+1.21}_{-1.31} & 0.31^{+0.06}_{-0.06} & 8.04^{+0.77}_{-0.75} & 0.20^{+0.04}_{-0.05} \\
SN~2006ef & NGC~0809 & 3.39^{+0.24}_{-0.27} & 0.61^{+0.07}_{-0.06} & 3.34^{+0.42}_{-0.36} & 0.54^{+0.05}_{-0.05} & 5.21^{+0.42}_{-0.96} & 0.28^{+0.03}_{-0.04} \\
SN~2006ej & IC~1563 & 4.79^{+2.30}_{-1.31} & 0.39^{+0.24}_{-0.22} & 8.13^{+4.07}_{-4.12} & 0.30^{+0.21}_{-0.23} & 8.40^{+3.04}_{-2.43} & 0.25^{+0.15}_{-0.22} \\
SN~2006kf & UGC~02829 & 9.27^{+0.84}_{-0.83} & 0.24^{+0.05}_{-0.05} & 16.52^{+1.13}_{-1.13} & 0.14^{+0.04}_{-0.05} & 15.40^{+1.13}_{-1.00} & 0.10^{+0.04}_{-0.04} \\
SN~2006ot & ESO~544-G031 & 4.83^{+1.58}_{-0.86} & 0.43^{+0.17}_{-0.16} & 8.40^{+2.72}_{-3.12} & 0.36^{+0.13}_{-0.15} & 8.82^{+2.12}_{-1.78} & 0.22^{+0.10}_{-0.12} \\
SN~2007R & MCG~+08-14-043 & 6.58^{+0.76}_{-0.67} & 0.24^{+0.06}_{-0.06} & 11.01^{+0.99}_{-0.85} & 0.21^{+0.05}_{-0.05} & 10.96^{+0.77}_{-0.82} & 0.11^{+0.04}_{-0.05} \\
SN~2007ap & MCG~+03-41-003 & 3.14^{+0.28}_{-0.25} & 0.61^{+0.07}_{-0.07} & 3.04^{+0.49}_{-0.27} & 0.56^{+0.06}_{-0.06} & 4.31^{+0.90}_{-0.54} & 0.31^{+0.05}_{-0.04} \\
SN~2007au & UGC~03725 & 7.36^{+0.56}_{-0.53} & 0.56^{+0.05}_{-0.04} & 16.61^{+0.89}_{-0.92} & 0.30^{+0.04}_{-0.04} & 14.74^{+0.79}_{-0.84} & 0.33^{+0.03}_{-0.04} \\
SN~2007ci & NGC~3873 & 4.82^{+0.40}_{-0.31} & 0.37^{+0.05}_{-0.05} & 7.89^{+0.87}_{-0.81} & 0.30^{+0.05}_{-0.04} & 8.34^{+0.50}_{-0.52} & 0.22^{+0.03}_{-0.04} \\
SN~2007cp & IC~0807 & 3.12^{+0.52}_{-0.40} & 0.37^{+0.13}_{-0.14} & 3.05^{+0.86}_{-0.57} & 0.40^{+0.10}_{-0.11} & 3.72^{+1.08}_{-0.52} & 0.21^{+0.09}_{-0.09} \\
SN~2007hj & NGC~7461 & 6.91^{+0.64}_{-0.55} & 0.30^{+0.05}_{-0.04} & 12.64^{+0.98}_{-0.94} & 0.16^{+0.04}_{-0.04} & 11.74^{+0.70}_{-0.67} & 0.18^{+0.04}_{-0.04} \\
SN~2007jh & CGCG~391-014 & 4.41^{+0.42}_{-0.37} & 0.34^{+0.07}_{-0.07} & 6.03^{+1.02}_{-0.79} & 0.33^{+0.06}_{-0.06} & 7.17^{+0.64}_{-0.60} & 0.19^{+0.04}_{-0.04} \\
SN~2007nq & UGC~00595 & 5.70^{+1.67}_{-1.10} & 0.42^{+0.14}_{-0.14} & 10.90^{+2.54}_{-2.69} & 0.27^{+0.13}_{-0.13} & 10.60^{+1.86}_{-1.86} & 0.23^{+0.09}_{-0.12} \\
SN~2008C & UGC~03611 & 0.10^{+\scalebox{0.6}{\textless}0.01}_{-\scalebox{0.6}{\textless}0.01} & 0.60^{+0.02}_{-0.02} & 0.64^{+0.05}_{-0.04} & 0.99^{+\scalebox{0.6}{\textless}0.01}_{-0.01} & 1.10^{+\scalebox{0.6}{\textless}0.01}_{-\scalebox{0.6}{\textless}0.01} & 0.07^{+0.17}_{-0.10} \\
SN~2008L & NGC~1259 & 3.23^{+0.36}_{-0.32} & 0.39^{+0.08}_{-0.09} & 3.30^{+0.62}_{-0.49} & 0.41^{+0.07}_{-0.07} & 4.15^{+0.71}_{-0.52} & 0.26^{+0.04}_{-0.07} \\
SN~2008R & NGC~1200 & 4.90^{+0.38}_{-0.27} & 0.61^{+0.06}_{-0.05} & 9.39^{+1.27}_{-1.18} & 0.43^{+0.04}_{-0.04} & 10.03^{+0.77}_{-0.78} & 0.30^{+0.04}_{-0.04} \\
SN~2008af & UGC~09640 & 3.88^{+0.37}_{-0.28} & 0.62^{+0.08}_{-0.08} & 4.93^{+1.37}_{-1.12} & 0.50^{+0.07}_{-0.07} & 6.88^{+0.85}_{-0.65} & 0.30^{+0.05}_{-0.05} \\
SN~2008gl & UGC~00881 & 3.11^{+0.46}_{-0.33} & 0.60^{+0.12}_{-0.10} & 3.14^{+0.78}_{-0.43} & 0.55^{+0.10}_{-0.09} & 4.63^{+0.90}_{-0.93} & 0.34^{+0.07}_{-0.07} \\
SN~2008hv & NGC~2765 & 4.62^{+0.28}_{-0.26} & 0.26^{+0.04}_{-0.04} & 6.07^{+0.68}_{-0.44} & 0.28^{+0.04}_{-0.03} & 7.26^{+0.46}_{-0.42} & 0.12^{+0.03}_{-0.03} \\
SN~2008ia & ESO~125-G006 & 4.55^{+0.16}_{-0.15} & 0.26^{+0.03}_{-0.03} & 5.88^{+0.30}_{-0.23} & 0.28^{+0.02}_{-0.02} & 7.05^{+0.25}_{-0.22} & 0.13^{+0.02}_{-0.02} \\
SN~2009F & NGC~1725 & 4.00^{+0.18}_{-0.16} & 0.76^{+0.05}_{-0.05} & 5.80^{+1.00}_{-0.94} & 0.55^{+0.04}_{-0.04} & 7.69^{+0.53}_{-0.56} & 0.40^{+0.04}_{-0.03} \\
SN~2010H & IC~0494 & 1.69^{+0.09}_{-0.17} & 0.75^{+0.09}_{-0.06} & 1.62^{+0.02}_{-0.03} & 0.67^{+0.04}_{-0.03} & 1.80^{+0.04}_{-\scalebox{0.6}{\textless}0.01} & 0.42^{+0.03}_{-0.03} \\
SN~2010Y & NGC~3392 & 3.65^{+0.36}_{-0.34} & 0.38^{+0.08}_{-0.09} & 4.15^{+0.80}_{-0.65} & 0.38^{+0.06}_{-0.07} & 5.24^{+0.61}_{-0.54} & 0.25^{+0.06}_{-0.06}\\
\enddata
\end{deluxetable*}
\clearpage

\begin{deluxetable}{lccc}
\tablewidth{0cm}
\tabletypesize{\scriptsize}
\centering
\caption{Slope and Significance of the Correlation with Hubble Residual \label{table6}}
\tablehead{\vspace{-8pt}\\
\colhead{} & \colhead{Age}  & \colhead{log(Age)} & \colhead{[M/H]}\\
\vspace{-20pt}\\
Model  & \colhead{($\Delta$mag/Gyr)} & \colhead{($\Delta$mag/Gyr)} & \colhead{($\Delta$mag/dex)}
}
\startdata
YEPS & -0.051 $\pm$ 0.022 & -0.67 $\pm$ 0.26 & 0.29 $\pm$ 0.26 \\
 & 98.98\% (2.3$\sigma$)  & 99.50\% (2.6$\sigma$)  & 86.77\% (1.1$\sigma$)  \\
TMJ11 & -0.021 $\pm$ 0.007 & -0.42 $\pm$ 0.14 & 0.70 $\pm$ 0.36 \\
 & 99.87\% (3.0$\sigma$)  & 99.87\% (3.0$\sigma$)   & 97.41\% (1.9$\sigma$)  \\
S07 & -0.027 $\pm$0.010 & -0.59 $\pm$ 0.21 & 0.45 $\pm$ 0.49 \\
 & 99.65\% (2.7$\sigma$)   & 99.75\% (2.8$\sigma$)   & 82.08\% (0.9$\sigma$) \\
\enddata
\end{deluxetable}

\clearpage

\begin{longrotatetable}
\begin{deluxetable*}{lllll}
\tablewidth{0cm}
\tabletypesize{\scriptsize}
\centering
\tablecaption{Correlation of SN Ia Brightness with Host Galaxy Property \label{table7}}
\tablehead{\colhead{Host Property} & \multicolumn{1}{l}{References} & \multicolumn{1}{l}{Original Correlation} &  \multicolumn{1}{l}{Direction} &  \colhead{Converted to Age Difference}
}
\startdata
Morphology & \citet{Hicken2009} & 
 $\Delta$HR/$\Delta$morph. 
 & Fainter in 
 & $\sim$0.19 mag/5.3 Gyr  \\
& & 
$\approx$0.14 mag/ (Scd/Irr---E/S0) 
& {\bf Later type} galaxy 
& Fainter in {\bf Younger} galaxy  \\
Mass & \citet{Sullivan2010} &
$\Delta$HR/$\Delta$mass 
& Fainter in 
& $\sim$0.21 mag/5.3 Gyr  \\
& & 
$\approx$0.08 mag/($\Delta$log $M_{\star} \sim$ 1 )
& {\bf Less massive} galaxy
& Fainter in {\bf Younger} galaxy \\
Local SFR & \citet{Rigault2018} &
$\Delta$HR/$\Delta$local SFR
& Fainter in 
& $\sim${0.34} mag/5.3 Gyr  \\
& & 
$\approx$0.16 mag/($\Delta$log LsSFR$_{\text{step}}$ $\sim$ 2  yr$^{-1}$ kpc$^{-2}$)
& {\bf Higher SFR} environments
& Fainter in {\bf Younger} galaxy \\
Population Age & This work & 
$\Delta$HR/$\Delta$age 
& Fainter in
& $\sim${0.27} mag/5.3 Gyr \\
& &
$\approx${0.051} mag/Gyr (YEPS) 
& {\bf Younger} galaxy 
& Fainter in {\bf Younger} galaxy \\
\enddata
\end{deluxetable*}
\end{longrotatetable}

\listofchanges
\end{document}